\documentclass[english,12pt,aps,prd,a4paper,preprintnumbers,floatfix,nofootinbib,showpacs,superscriptaddress, notitlepage]{revtex4-1} 
 \pdfoutput=1
\usepackage[usenames,dvipsnames]{color}  %%%%%%% usenames,dvipsnames required to get BrickRed
\usepackage{graphicx}
\usepackage{setspace}
\usepackage{caption}
\usepackage{subcaption}
\usepackage{amsmath}
\usepackage{amssymb}
\usepackage[colorlinks=true,citecolor=darkred,urlcolor=darkred, pdfborder={0 0 0}]{hyperref}
\usepackage[normalem]{ulem}
\usepackage{xcolor}

\makeatletter
\def\p@subsection{}
\makeatother

\definecolor{darkred}{rgb}{0.6,0,0}

\definecolor{linkcolor}{rgb}{0,0,0.5}

\def\gsim{\raise0.3ex\hbox{$\;>$\kern-0.75em\raise-1.1ex\hbox{$\sim\;$}}}
\def\lsim{\raise0.3ex\hbox{$\;<$\kern-0.75em\raise-1.1ex\hbox{$\sim\;$}}}

\def\beqn#1{\begin{equation}\label{#1}}
\def\eeqn{\end{equation}}

\def\beqa#1{\begin{eqnarray}\label{#1}}
\def\eeqa{\end{eqnarray}}

\def\znbb {neutrinoless double beta decay }

\def\Z2{$\mathcal{Z_2}$}

\def\vev#1{\left\langle #1\right\rangle}

\newcommand {\ignore}[1]{}

\newcommand{\sm}{{Standard Model }}

\def\SM{$\mathrm{SU(3)_c \otimes SU(2)_L \otimes U(1)_Y}$ }

\def\321{$\mathrm{SU(3) \otimes SU(2) \otimes U(1)}$ }

\newcommand{\AddrAHEP}{
  AHEP Group, Institut de F\'{i}sica Corpuscular --
  CSIC/Universitat de Val\`{e}ncia, Parc Cient\'ific de Paterna.\\
 C/ Catedr\'atico Jos\'e Beltr\'an, 2 E-46980 Paterna (Valencia) - SPAIN}
 
\newcommand{\AddrIISERB}{Department of Physics, Indian Institute of Science Education and Research - Bhopal, \\ 
Bhopal Bypass Road, Bhauri, Bhopal 462066, India}

\begin{document}

 \bibliographystyle{unsrt}   % needed for refs and hyperlinks %% utphys.bst style file is also needed with this %%%%%%%%%%%%
 
\title{Consistency of the dynamical high-scale type-I seesaw mechanism}

\author{Sanjoy Mandal}\email{smandal@ific.uv.es}
\affiliation{\AddrAHEP}
\author{Rahul Srivastava}\email{rahul@iiserb.ac.in}
\affiliation{\AddrIISERB}
\author{Jos\'{e} W. F. Valle}\email{valle@ific.uv.es}
\affiliation{\AddrAHEP}

\begin{abstract}
 \vspace{1.2 cm}
 
We analyze the consistency of electroweak breaking within the simplest high-scale \SM type-I seesaw mechanism. We derive the full two-loop RGEs of the relevant
parameters, including the quartic Higgs self-coupling of the Standard Model.  
For the simplest case of bare ``right-handed'' neutrino mass terms we find that, with large Yukawa couplings, the Higgs quartic self-coupling becomes 
negative much below the seesaw scale, so that the model may be inconsistent even as an effective theory. 
We show, however, that the ``dynamical'' type-I high-scale seesaw with spontaneous lepton number violation has better stability properties.

\end{abstract}

\maketitle
%%%%%%%%%%%%%%%%%%%%%%%%%%%%%%%%%%%%%%%%%%%%%%%%%%%%%%%%%%%%%%%%%%%%

\section{Introduction}
\label{sec:introduction}

%%%%%%%%%%%%%%%%%%%%%%%%%%%%%%%%%%%%%%%%%%%%%%%%%%%%%%%%%%%%%%%%%%%

The discovery of a scalar particle with 125 GeV mass plays a central role within particle physics~\cite{Aad:2012tfa,Chatrchyan:2012xdj}. In particular, the precise Higgs boson mass measurement determines the value of the quartic coupling in the scalar potential at the electroweak scale and allows one to study its behavior all the way up to high energies. 
Given the measured values of \sm parameters such as the top quark and Higgs boson masses, we know that the Higgs quartic coupling remains perturbative after renomalization group equations (RGEs) are used to evolve it to high energies. However, the stability of the fundamental vacuum may fail at mass scales below the fundamental Planck scale~\cite{Tanabashi:2018oca}. 

Another most important milestone in particle physics has been the discovery of neutrino oscillations~\cite{Kajita:2016cak,McDonald:2016ixn}. 
This implies the existence of neutrino masses~\cite{deSalas:2017kay} and hence new physics that can produce them~\cite{Valle:2015pba}. 
Electroweak vacuum stability can be substantially affected in the presence of a dynamical seesaw mechanism~\cite{Lindner:2015qva,Bambhaniya:2016rbb}
~\footnote{Here we will focus on stability within high-scale seesaw. For discussions of low-scale seesaw see~\cite{Bonilla:2015kna,Rose:2015fua}.}.

Here we examine more closely the issue of the consistency of the Higgs vacuum within type-I seesaw extensions of the \SM Standard Model with ungauged lepton number~\cite{Schechter:1980gr}.
For ``sizeable'' Yukawa coupling, $Y_{\nu}\sim\mathcal{O}(1)$, in order to reproduce the required neutrino masses, heavy neutrinos must lie at mass scale $M_{N} \sim \mathcal{O}(10^{14}\,\text{GeV})$.
This characterizes the case of genuine ``high-scale'' type-I seesaw constructions.
We stress that \SM seesaw extensions can be formulated with any number of ``right-handed'' neutrinos, since they carry no anomaly. 
Here for definiteness, we start from the minimalistic (3,1) model containing only one right-handed neutrino, in addition to the 3 known left-handed neutrinos~\cite{Schechter:1980gr}.
We start from this ``missing partner'' seesaw, aware of, by itself, it does not provide a fully realistic picture, since only one neutrino mass scale arises at the tree level~\cite{Schechter:1980gr}.
However, there are interesting and realistic variants where the ``missing partner'' seesaw scale can be identified with the ``atmospheric scale'', while
the ``solar'' mass scale is generated by radiative corrections that could arise, for example, from a ``dark matter'' sector~\cite{Rojas:2018wym}. 
We therefore take such ``missing partner'' seesaw as our reference scheme. 

We show that, although it has better stability properties than the fully sequential (3,3) seesaw mechanism, for sizeable magnitudes of the Yukawa couplings the Higgs potential of the minimal (3,1) seesaw becomes unstable even below the seesaw scale.
The situation can only get worse by having more right-handed neutrinos, in the (3,2) seesaw or in the ``sequential'' (3,3) seesaw.
In this sense too it makes sense to take such ``missing partner'' (3,1) seesaw as the reference scheme.
An important implication of this missing partner type-I seesaw is the existence of a lower bound on the neutrinoless double beta decay rate even for normal ordered neutrino masses~\cite{Valle:2020wdf}. 

We then show explicitly that vacuum stability can be improved naturally if one implements spontaneous violation of lepton number. This is characterized by the existence of a physical Nambu-Goldstone boson, dubbed majoron~\cite{Chikashige:1980ui,Schechter:1981cv}. 
We show how the extra scalars required to implement spontaneous lepton number violation play a key role to improve stability properties.
Indeed, their couplings can easily restore stability of the electroweak symmetry breaking even if the lepton number violation scale is high, as required to fit neutrino masses in this case. 
We also analyse the scale at which instability sets in as a function of the magnitude of the Yukawa coupling relevant for generating neutrino mass in (3,1), as well as as the conventional (3,3) seesaw case.

This work is organized as follows: In Section~\ref{sec:sm-vac} we revisit the vacuum stability problem in the \sm showing that the Higgs quartic coupling becomes negative when RGEs-evolved to high scales. 
In Section~\ref{sec:neutr-mass-gener}, we describe the neutrino mass generation in type-I seesaw and type-I seesaw with majoron extensions. 
We then show in Section~\ref{Higgs vacuum stability and neutrino mass} that the vacuum stability problem becomes worse in high-scale type-I seesaw Standard Model extensions.
We then focus on the majoron extension of the canonical type-I seesaw. We show how the majoron helps stabilize the Higgs vacuum, which can be made stable all the way up to Planck scale. 
In addition the majoron could provide a viable dark matter candidate~\cite{Berezinsky:1993fm,Lattanzi:2007ux,Bazzocchi:2008fh,Lattanzi:2013uza,Kuo:2018fgw}, thereby solving another basic problem in particle physics.
Finally, we conclude and summarize our main results in Section~\ref{sec:summary-discussion}.

%%%%%%%%%%%%%%%%%%%%%%%%%%%%%%%%%%%%%%%%%%%%%%%%%%%%%%%%%%%%%%%%%%%%%%%

\section{Higgs vacuum in the \sm}
\label{sec:sm-vac}

%%%%%%%%%%%%%%%%%%%%%%%%%%%%%%%%%%%%%%%%%%%%%%%%%%%%%%%%%%%%%%%%%%%%%%%

Let's briefly revisit the status of the electroweak (EW) vacuum within the Standard Model. 
For a long time the Higgs boson was the ``last'' missing piece of the theory. 
The discovery of a scalar particle with mass $m_H \approx 125$ GeV at the Large Hadron Collider (LHC) is very suggestive that it could be the long-awaited \sm Higgs boson.
While further work is still needed to unambiguously establish this, current data indicates that its couplings and decay properties are close to the \sm Higgs expectations.
If, indeed, this is the case, the next question is, given that so far we have not seen any evidence for new particles at the LHC, whether the \sm can be the final theory.
The answer is obviously no, since the \sm predicts neutrinos to be massless and there is no viable \sm candidate for cosmological dark matter.

For the moment we put these two issues aside, and ask ourselves whether there are other compelling hints that the \sm cannot be the final theory up to Planck scale.  
Indeed, there are several other theoretical and aesthetical arguments against this being the case. For example, achieving the unification of forces and the improving the hierarchy/fine-tunning/naturalness problem.
However, the ``Higgs discovery'' has facilitated us to study the high energy behavior of the Standard Model. 
As an example, in this work we address the stability of Higgs vacuum at energies far above the electroweak scale.

The detailed analysis of the Higgs vacuum within the \sm has been carried out in~\cite{Isidori:2001bm,EliasMiro:2011aa,Bezrukov:2012sa,Degrassi:2012ry,Masina:2012tz,Buttazzo:2013uya}.
For completeness here we revisit this analysis. This serves us to calibrate our Renormalization Group analysis against known results.
Although in dedicated \sm studies there are some partial 3-loop results~\cite{Buttazzo:2013uya}, 
to compare the seesaw and \sm results it will suffice for us to stay at the two-loop level.
In our analysis we adopt the $\overline{MS}$ scheme, taking the parameter values at low scale as the input values~\cite{Tanabashi:2018oca}. 
In particular, the Higgs pole mass is taken as the current best fit value of $m_H = 125.18\pm 0.16$ GeV, the top quark pole mass is taken as $m_t = 173\pm 0.4$ GeV and 
the strong coupling constant $\alpha_{s}(M_{Z})=0.1184\pm 0.0007$.
Using these experimental values, we adopt the ``On-Shell'' renormalization scheme in order to express the renormalized parameters directly in terms of the physical 
observables and then relate the on-shell parameters to the $\overline{MS}$ parameters in a way similar to \cite{Buttazzo:2013uya}. 
In Table.~\ref{input parameter} we list the $\overline{MS}$ input values of the relevant parameters at the top mass $m_t$ scale. 
\begin{table*}[ht]
	\centering
	\begin{tabular}{|c|c|c|c|c|c|}
		\hline
		                 &   $g_{1}$  &  $g_{2}$  &  $g_{3}$  & $y_{t}$  &  $\lambda_{\text{SM}}$  \\
		\hline
		$\mu (m_t)$      &   0.462607                   &  0.647737 &  1.16541  & 0.93519 &  0.126115   \\
		\hline
	\end{tabular}
	\caption{\footnotesize{$\overline{MS}$ values of the main input parameters at the top quark mass scale, $m_{t}=173\pm 0.4$ GeV. }}
	\label{input parameter}
\end{table*}  

\begin{figure}[h]
\centering
\includegraphics[width=0.75\textwidth]{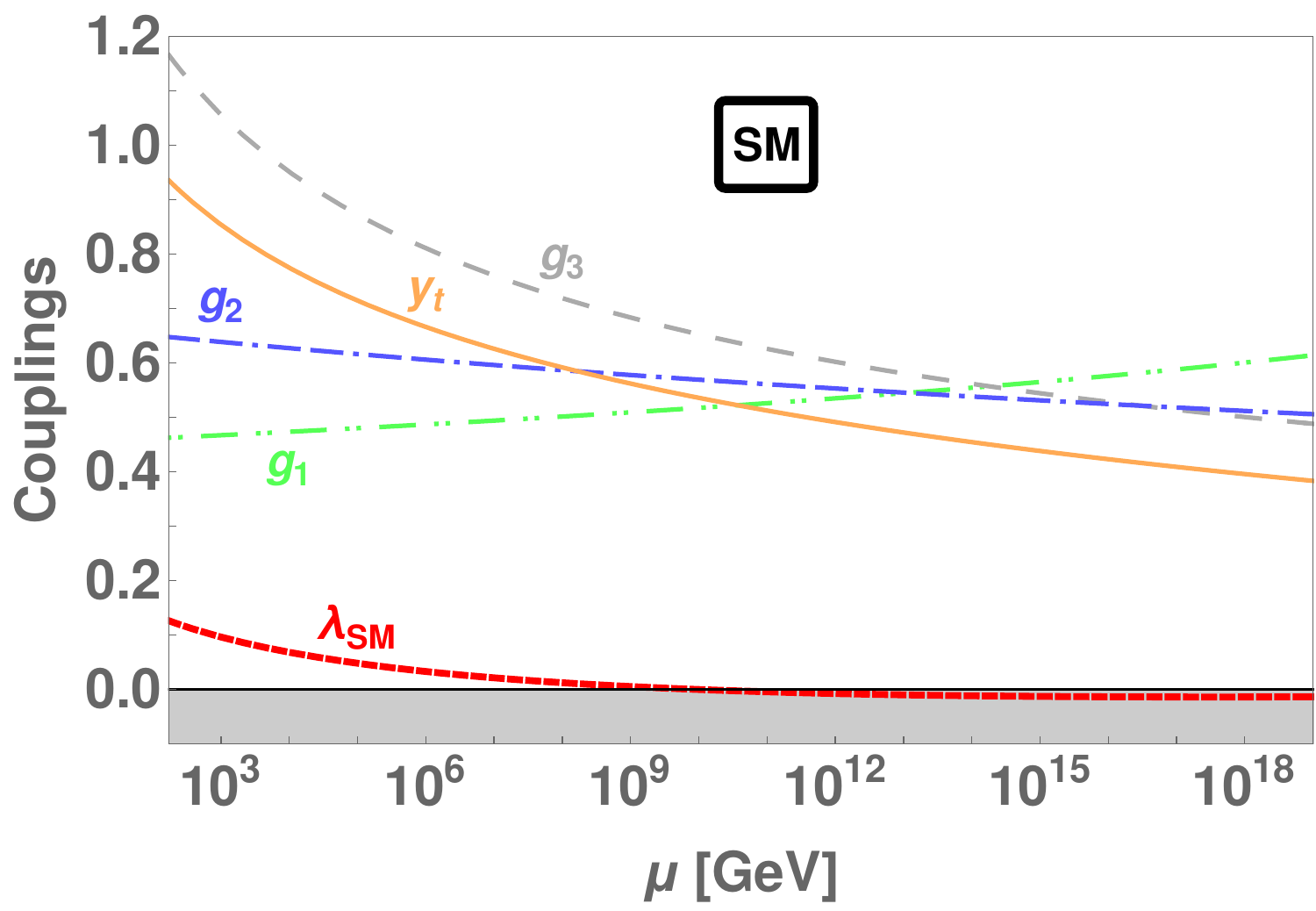}
\caption{\footnotesize{The RG evolution of the \sm gauge couplings $g_{1}$, $g_{2}$, $g_{3}$, 
the top quark Yukawa coupling $y_{t}$ and the quartic Higgs boson self-coupling $\lambda_{\text{SM}}$.}} 
\label{RG-SM-Running}
\end{figure}

Taking the initial $\overline{MS}$ values of Table.~\ref{input parameter} as input values, we then RGEs-evolve the \sm parameters to higher scales as shown in Fig.~\ref{RG-SM-Running}.  

Taking into account the updated input parameter values, our two-loop results are in good agreement with earlier ones. 
Tiny differences arise mainly due to the increased current precision of the experimental numbers.
We stress that an in-depth reanalysis of the \sm Higgs is not the goal of our paper, but rather the comparison of \sm and seesaw scenarios.
Hence, we refrain from performing a sensitivity analysis of Higgs vacuum stability and its dependence on the input parameter errors. 
Indeed, in the seesaw scenarios of interest to us, such tiny effects are negligible when compared to the effects of the new Yukawa couplings. 

Notice from Fig.~\ref{RG-SM-Running} that the \sm Higgs quartic coupling $\lambda_{\text{SM}}$ becomes negative at $\mu \simeq 10^{10}$ GeV. 
This would imply that the Higgs potential is unbounded from below and the Higgs vacuum would be unstable.  
A dedicated analysis shows that, in fact, the \sm Higgs vacuum is not unstable, but rather metastable~\footnote{\sm vacuum stability is sensitive to input parameter values, in particular the top-quark mass.} with very long lifetime~\cite{Buttazzo:2013uya}. 

\section{Neutrino mass generation}
\label{sec:neutr-mass-gener}

As already mentioned, the \sm cannot be the final theory up to the Planck scale, as it has massless neutrinos and no viable candidate for dark matter.
Hence the vacuum stability issue must be re-considered. % in \sm extensions with the potential to explainhich can address these issues.
%Hence the vacuum stability analysis must be carried out within Standard Model extensions which can address these issues.
%
We now do this adopting simple seesaw extensions of the Standard Model.
We show that, above the seesaw scale, the Higgs vacuum stability can be completely dominated by the new couplings.
Hence it suffices for our purposes to discuss electroweak vacuum stability at the two-loop level. %s, without going into fine details about input
%parameters~\footnote{It is straightforward to refine the seesaw analysis along the lines presented in \cite{Buttazzo:2013uya} for the \sm.}

%%%%%%%%%%%%%%%%%%%%%%%%%%%%%%%%%%%%%%%%%%%%%%%%%%%%%%%%%%%%%%%%%%%%
\subsection{Dimension-five operator}
\label{sec:dim5}
%%%%%%%%%%%%%%%%%%%%%%%%%%%%%%%%%%%

Within the \sm neutrinos are massless. However, as first noted by Weinberg~\cite{Weinberg:1979sa}, non-zero masses will arise from an effective non-renormalizable 
dimension-five operator characterizing lepton number non-conservation.
The effective Lagrangian reads

\begin{align}
 -\mathcal{L}_{\nu}^{d=5}=\frac{1}{2}\left(\bar{\ell}_{L}H\right).\kappa . \left(H^{T}\ell_{L}^{c}\right)+h.c.~,
\end{align}
where $\kappa$ is the $3 \times 3$ symmetric coupling matrix with negative mass dimension and, for brevity, we have suppressed the generation indices.
When the electroweak symmetry breaking occurs, the Higgs gets vacuum expectation value (vev) $\vev{H} = \frac{v}{\sqrt{2}}$ with $v = 246\,\text{GeV}$,
$H$ being the \sm Higgs doublet. 
This leads to a Majorana mass matrix for the left-handed neutrinos given as
%After the electroweak symmetry breaking, the Higgs gets vacuum expectation value (vev) $\vev{H} = \frac{v}{\sqrt{2}}$ with $v = 246\,\text{GeV}$.
%
$$m_{\nu}\equiv \kappa \frac{v^{2}}{2}.$$
which leads to light neutrino masses and lepton number violation by two units.
There are many ways to generate $\kappa$ as a result of postulating new mediator particles. 
A very simple ``UV-completion'' is the type-I seesaw mechanism.

\subsection{Type-I seesaw mechanism }
\label{sec:seesaw}

%%%%%%%%%%%%%%%%%%%%%%%%%%%%%%%%%%%%%%%%%%%%%%%%%%%%%%%%%%%%%%%%%%%

The most general ``type-I seesaw'' mechanism is the one formulated in terms of just the \SM structure characterizing the Standard Model,
without extra gauge symmetry~\cite{Schechter:1980gr}.
One postulates the existence of gauge singlet  ``right-handed neutrinos'', $\nu_{R_i}$, $i = 1,2,\cdots n$, whose mass term is obviously gauge invariant.
Neutrino masses arise from the exchange of ``right-handed neutrino'' mediators whose multiplicity is arbitrary since, as gauge singlets, they carry no anomaly. 
The relevant part of the Lagrangian is written as
\begin{align}
- \mathcal{L} = \sum_{a,i}Y_{\nu}^{ai} \bar{\ell}_{L}^{a} \tilde{H} \nu_{R_i} + \frac{1}{2}\sum_{i,j}M^{ij}_{R}\overline{\nu_{R_i}^{c}}\nu_{R_j} + \text{H.c.}
\label{seesaw-lag}
\end{align}  
where $\ell_{L}^{a} = (\nu^a_L, l^a_L)^T$ with $a=1,\,2,\,3$ denotes the three families of left-handed lepton doublets, while $i,j=1,2,\cdots n$ labels the right-handed 
singlet neutrinos and, as before, $H$ is the \sm Higgs doublet. 
After electroweak symmetry breaking the full neutrino mass matrix is expressed as
\begin{align}
\mathcal{M}_{\nu}=
\begin{pmatrix}
 0 & m_{D} \\
 m_{D}^{T} & M_{R}  \\
\end{pmatrix}
\label{mass-matrix}
\end{align}
where $m_{D}=\frac{Y_{\nu}}{\sqrt{2}}v$ is the ``Dirac mass matrix''. 

Being \SM invariant, the right-handed neutrino ``Majorana mass matrix'' $M^{ij}_R$ entries can be much larger than the EW scale, $|M^{ij}_R| >> v$, 
implying  $|\frac{m^{aj}_{D}}{M^{ij}_{R}}| << 1$. 
Hence the mass matrix in Eq.\eqref{mass-matrix} can be block-diagonalized perturbatively in an exponential series, Eq.(3.1) in~\cite{Schechter:1981cv}\footnote{
We are now using the series expansion in Ref.~\cite{Schechter:1981cv} for the case of explicit lepton number violation.}. 
The two diagonalized blocks correspond to ``light'' and ``heavy'' neutrino mass matrices, denoted as $m^{ab}_{\nu}$ and $M^{ab}_{N}$ respectively, which can be written symbolically as:
\begin{eqnarray}
 m^{ab}_\nu = \frac{1}{2}\left [ M^{ab}_R - \left (\sqrt{M^2_R + 4 m^2_D} \right )^{ab} \right ]; \quad
 M^{ab}_N = \frac{1}{2}\left [ M^{ab}_R + \left (\sqrt{M^2_R + 4 m^2_D} \right )^{ab} \right ]
 \label{block-form}
\end{eqnarray}
%
%
% Note that for (3,1) type-I seesaw, light and heavy neutrino mass eigenvalues are given by,
% \begin{align*}
%  m_{\nu_1}=\frac{1}{2}(M_R-\sqrt{M_R^2+4m_D^2}),\,\,M_N=\frac{1}{2}(M_R+\sqrt{M_R^2+4m_D^2})
% \end{align*}
% Under the approximation $M_R\gg m_D$, we get the standard simple type-I seesaw formulae $m_{\nu_1}\approx \frac{m_D^2}{M_R}$ and $M_N\approx M_R + \frac{m_D^2}{M_R}$.}
%
To leading order the mass matrix elements for light neutrinos $m^{ab}_{\nu}$ and heavy neutrinos $M^{ab}_{N}$, are given as
\begin{eqnarray}
m^{ab}_{\nu} & \simeq &  -m^{ai}_{D} (M_{R}^{-1})^{ij}(m_{D}^{T})^{jb} \, + \, \text{higher order terms}, \label{seesaw} \\
M^{ab}_{N} & \simeq &  M^{ab}_R \, +\, m^{ai}_{D} (M_{R}^{-1})^{ij}(m_{D}^{T})^{jb} \, + \, \text{higher order terms}, 
\label{heavy-seesaw}
\end{eqnarray}
where the negative sign in \eqref{seesaw} can be absorbed~\footnote{Note that, even though this negative sign in \eqref{seesaw} is physical, related to the CP properties of the light neutrinos,
  for our purposes they will not be relevant. } through field redefinition.
The full expression for the diagonalizing matrix is found in~\cite{Schechter:1981cv}, as Eq.(3.5). 
The light neutrino mass matrix in \eqref{seesaw} is further diagonalized by a unitary matrix $U_\nu$ in the light neutrino sector $\nu^a$; $a = 1,2,3$.
This famous type-I seesaw formula links the smallness of the light neutrino masses to the heaviness of the right-handed neutrinos $\nu_R$. 

\subsection{ The missing partner type-I seesaw mechanism }
\label{sec:miss-seesaw}

As already mentioned, since $\nu_R$'s are \sm gauge singlets, their number $n$ need not match the number of left-handed ones.
Depending on the value of ``$n$'' many possibilities can be envisaged. Here we consider the case $n\leq3$ of ``high-scale'' constructions\footnote{Here 
we discard cases with $n>3$ since having extra fermions can only worsen stability. 
An interesting example would be the (3,6) seesaw scheme, which includes the template for the sequential ``low-scale'' seesaw, including both the inverse 
seesaw~\cite{Mohapatra:1986bd,GonzalezGarcia:1988rw} and the linear seesaw mechanisms~\cite{Akhmedov:1995ip,Akhmedov:1995vm,Malinsky:2005bi}.}.

The observation of neutrino oscillations~\cite{Kajita:2016cak,McDonald:2016ixn} proves that two of the three ``active'' neutrinos are massive~\cite{deSalas:2017kay}.
However, there is so far no indication for a finite mass for the lightest neutrino.
Indeed, the Katrin experiment has derived an upper limit of 1.1 eV (at 90\% C.L.) on the absolute mass scale of 
neutrinos~\cite{Aker:2019uuj} from the Tritium endpoint spectrum. 
This bound applies irrespective of whether neutrinos are Dirac or Majorana particles.
On the other hand, cosmological observations indicate that $\sum m_a \leq 0.12$ eV~\cite{Aghanim:2018eyx,Lattanzi:2017ubx}. 
Hence, for ``sizeable'' Yukawa coupling values, $Y_{\nu}\sim\mathcal{O}(1)$, this bound is satisfied for heavy neutrino masses $M_{N}\sim \mathcal{O}(10^{14}\,\text{GeV})$.

In order to account for the current oscillation evidence for neutrino mass
it suffices to have a ``missing partner'' seesaw mechanism with $n=2$, since in this case both solar and atmospheric scales can be produced by the seesaw. 
Following the general formulation in~\cite{Schechter:1980gr} we call such scheme in which one left-neutrino has no right partner, (3,2) seesaw.
In this case each right-handed neutrino mediates the generation of the corresponding scale, solar or atmospheric.

Notice, however, that the minimal type-I seesaw mechanism is the one in which only one right-handed neutrino is added to the \sm.
This (3,1) scheme is the minimal ``missing partner'' seesaw, in which two left-neutrinos lack a right-partner and remain massless. 
It can be viable as part of a bigger scheme in which, for example, the solar scale arises radiatively, hence accounting the small solar/atmospheric scale ratio.
Such scheme is easily obtained by ``cloning'' the seesaw with some other sector associated, for example, with dark matter.
An interesting realization is the scoto-seesaw, in which the atmospheric scale is seesaw-induced, while the solar scale has scotogenic origin~\cite{Rojas:2018wym}.

\begin{figure}[h]
    \centering
\includegraphics[height=7cm,width=0.8\textwidth]{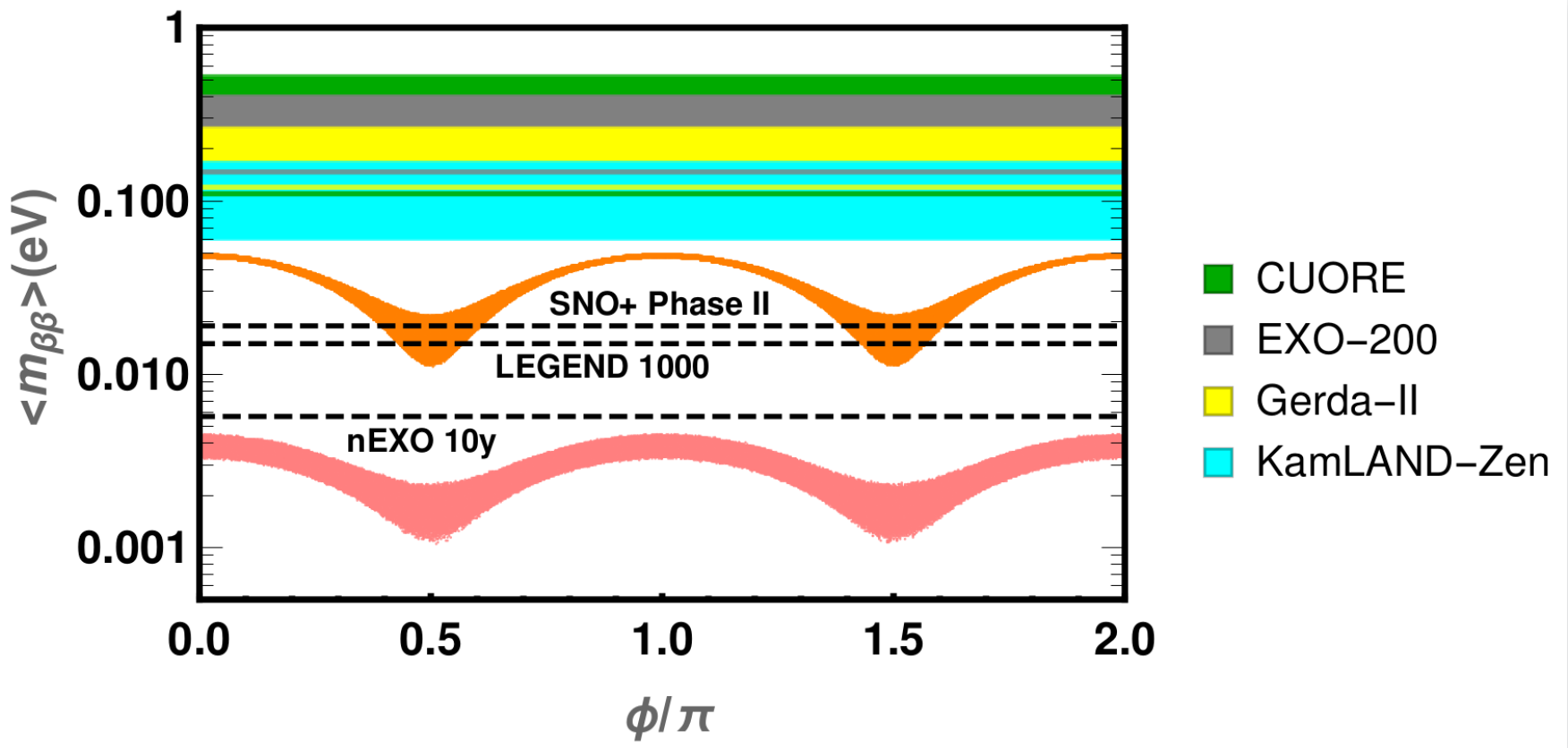}~~~~~
\caption{Amplitude for \znbb in missing partner seesaw. The horizontal bands represent the limits from current experiments, while the horizontal dashed lines show the maximum reach of future experiments (see text). }
    \label{fig:dbd}
  \end{figure}

An implication of the missing partner seesaw schemes is a prediction for the parameter $m_{\beta\beta}$ describing the amplitude 
for \znbb versus the (relative) massive neutrino Majorana phase, shown in Fig.~\ref{fig:dbd}

The lower band corresponds to normal mass ordering and the upper one to inverted. 
Their narrow widths reflect the small allowed spread in neutrino oscillation parameters~\cite{deSalas:2017kay}. 
Notice that, in contrast to the general ``complete'' (3,3) seesaw, in the missing partner there can be no cancellation, so that
non-zero \znbb is predicted, even if neutrinos are normal-ordered.
The horizontal bands in Fig.~\ref{fig:dbd} show the reach of present experiments:
  CUORE (green, limits: 0.11 - 0.52 eV)~\cite{Alduino:2017ehq}, EXO-200 (grey, limits: 0.147 - 0.398 eV)~\cite{Albert:2017owj}, Gerda-II (yellow, limits: 0.120 - 0.260 eV)~\cite{Agostini:2018tnm}
  and KamLAND-Zen (cyan, limits: 0.061 - 0.165 eV)~\cite{KamLAND-Zen:2016pfg}. The horizontal lines indicate the maximum estimated experimental sensitivities~\footnote{Here we made the most optimistic assumptions concerning nuclear matrix element uncertainties.} of upcoming experiments: SNO+ Phase-II (0.019 eV)~\cite{Andringa:2015tza}, LEGEND - 1000 (0.015)~\cite{Abgrall:2017syy} and nEXO - 10yr (0.0057)~\cite{Albert:2017hjq}.
One sees from Fig.~\ref{fig:dbd} that, although the upcoming experiments are only sensitive to inverted ordering, the
 detectability chances improve in the ``missing partner'' as compared to the expectations of a generic ``complete'' seesaw mechanism.
This brings hope that upcoming experiments may be able to measure, for the first time, the relevant Majorana phase.\\[-.3cm]

To sum up, in what follows we take the missing partner seesaw as our reference benchmark, because of its minimality and notational simplicity, and also because of the fact that having extra fermions can only worsen stability of the Higgs potential.
In addition, in the ``complete'' seesaw picture one looses the \znbb prediction in Fig.~\ref{fig:dbd}.

However, in Sec.~\ref{sec:3gen} we explicitly compare our results with those obtained for the sequential (3,3) seesaw.
Moreover, all the relevant renormalization group equations given in the Appendix assume the conventional (3,3) seesaw picture.

%%%%%%%%%%%%%%%%%%%%%%%%%%%%%%%%%%%%%%%%%%%%%%%%%%%%%%%%%%%%%%%%%%
\section{Higgs vacuum stability and neutrino mass}
\label{Higgs vacuum stability and neutrino mass}
%%%%%%%%%%%%%%%%%%%%%%%%%%%%%%%%%%%%%%%%%%%%%%%%%%%%%%%%%%%%%%%%%

We saw how the \sm vacuum is not absolutely stable. Instead, with the present measured values of Higgs and top masses, it is metastable, as the quartic coupling $\lambda_{\text{SM}}$ becomes negative around $\Lambda_{I}^{SM}\sim 10^{10}$~GeV, the \sm instability scale.
Before discussing Higgs vacuum stability within type-I seesaw embeddings of the Standard Model we first consider the effective Weinberg operator.

%%%%%%%%%%%%%%%%%%%%%%%%%%%%%%%%%%%%%
\subsection{Dimension-five operator}
\label{sec:dim5a}
%%%%%%%%%%%%%%%%%%%%%%%%%%%%%%%%%%%%%%%%%

\begin{figure}[h]
\centering
\includegraphics[height=3.5cm,width=0.45\textwidth]{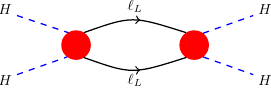}
\caption{\footnotesize{Effect of Weinberg's effective operator on the Higgs quartic interaction in the effective theory.} }
\label{dimension five operator}
\end{figure}

We now discuss the running of the quartic scalar coupling characterizing the EW symmetry breaking sector of the \sm in the context of a 
dimension-five operator picture that results effectively at low-energies from a UV-complete type-I seesaw.
Below the scale $\mu=M_R$ we integrate out the heavy neutrino $\nu_R$, so that the theory is the \sm
plus an effective dimension-five Weinberg operator $\mathcal{L}_{\nu}^{d=5}$~(with $\kappa=Y_{\nu}M_{R}^{-1}Y_{\nu}^{T}$). 
Below the scale $\mu=M_R$, only the \sm couplings and $\kappa$ will run. 
Neglecting the contribution from lepton and light quark Yukawa couplings, the one-loop RGEs
is given by~\cite{Chankowski:2000fp,Antusch:2002rr,Bergstrom:2010id} (see also Ref.~\cite{Bonilla:2015kna}),
\begin{align}
 16\pi^{2}\beta_{\kappa}=6y_{t}^{2}\, \kappa-3g_{2}^{2} \,\kappa + \lambda_\kappa \, \kappa.
\end{align}
As the SM case, here $y_t$ also denotes the top Yukawa coupling and $g_2$ is the $SU(2)_L$ gauge coupling.
  We denote the Higgs quartic coupling in this case as $\lambda_\kappa$ to distinguish it from the pure \sm case.
Hence, due to the large top Yukawa coupling $y_{t}$, the coupling $\kappa$ slowly increases with the energy scale $\mu$.
As seen in Fig.~\ref{dimension five operator}~\cite{Bonilla:2015kna}, the same operator which generates the neutrino 
mass below the scale $\mu=M_R$, also provides a correction to the Higgs quartic self-coupling $\lambda_\kappa$ below that scale.
The contribution from the coupling $\kappa$ on the running of the Higgs quartic coupling $\lambda_\kappa$ is of order $v^{2}\kappa^{2}$ 
and thus negligible~\cite{Ng:2015eia}. 
As a result, in the effective theory, the running of $\lambda_\kappa$ below the scale $\mu=M_{R}$ will be almost the same as in
the Standard Model. 

%%%%%%%%%%%%%%%%%%%%%%%%%%%%%%%%%%%%%%%%%%%%%%%%%%%%%%%%%%%%%%%%%
\subsection{Higgs Vacuum Stability in Type-I Seesaw} 
\label{sec:result-seesaw}
%%%%%%%%%%%%%%%%%%%%%%%%%%%%%%%%%%%%%%%%%%%%%%%%%%%%%%%%%%%%%%%%%%

\begin{figure}[h]
\centering
\includegraphics[height=4cm,width=0.3\textwidth]{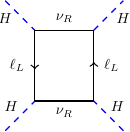}
\caption{\footnotesize{One-loop correction to the Higgs quartic interaction in the full seesaw theory.} }
\label{Loop Correction to lambda in full theory}
\end{figure}

In what follows we will be mostly concerned with the effects of sizable Yukawa couplings in the context of high-scale 
missing partner seesaw and their impact on the stability of the Higgs vacuum.
Building upon the discussion of the previous section, we now turn to the region above the scale $\mu=M_R$. 
In this case one has the full theory in which the running of Yukawa coupling $Y_{\nu}$ will have an impact on the running of the Higgs quartic coupling which we now call $\lambda$. This is done so as to distinguish it from both the \sm case as well as from the regime where the renormalization group running is performed only with the effective Weinberg operator. 
Within the type-I seesaw picture, below and above the seesaw scale $\mu=M_R$, there will be contributions on Higgs quartic coupling $\lambda$ from
the Figs.~\ref{dimension five operator} and \ref{Loop Correction to lambda in full theory}, respectively. Hence, in order to describe the running of $\lambda$
we need to take into acount the matching condition at the scale $\mu=M_R$.

For the reasons mentioned in Sec.~\ref{sec:seesaw} here we focus on the simplest missing partner type-I seesaw mechanism containing a single right-handed neutrino.
It provides a clear picture of the impact of seesaw extensions on the Higgs vacuum stability in the simplest possible setting.

As we discussed earlier, below the $\mu=M_{R}$ scale, the theory is an effective \sm supplemented by the dimension-five Weinberg operator.
However, above the $\mu=M_{R}$ scale the theory is UV-complete, so that all the new couplings in the model like the neutrino Yukawa 
coupling $Y_{\nu}$ will take part in the system of renormalization group equations and will affect the running of the \sm couplings, 
specially that of $\lambda$. 

As a result, the stability of the electroweak vacuum will set a potential limit on how large $Y_{\nu}$ can be.
As the new Yukawa coupling $Y_{\nu}$ runs only above the threshold scale $M_{R}$, this can be technically implemented 
by replacing $Y_{\nu}\to Y_{\nu}\theta\left(\mu - M_{R}\right)$  in the right hand side of the RGEs of the full theory, given in Appendix~\ref{app:seesaw}. Here $\theta(x) = 1, x>0 $ and 
$\theta(x) = 0, x<0 $ is the step function.

Integrating out the heavy neutrinos also introduces threshold corrections to the \sm Higgs quartic coupling
$\lambda$~\cite{Casas:1999cd} at the scale $\mu=M_R$. The tree level Higgs potential in the \sm
is given by
\begin{align}
 V\left(H^{\dagger}H\right)=-\mu_{H}H^{\dagger}H+\lambda \left(H^{\dagger}H\right)^{2}.
\end{align}
This will get corrections from higher loop diagrams of \sm particles and extra fermion from the type-I seesaw. 
The one-loop effective potential $V_{1}(h)$ has the form
\begin{align}
 V_{1}(h)=V_{1}^{\text{SM}}(h)+V_{1}^{\nu}(h)
\end{align}
where $V_{1}^{\text{SM}}(h)$ is the usual one loop \sm potential. % and can be found in Ref.~\cite{Quiros:1997vk}. 
The one loop potential from the neutrino sector is given by~\cite{Brivio:2017dfq,Casas:1999cd}
\begin{align}
V_{1}^{\nu}(h)=-\frac{1}{32\pi^{2}}\left(\sum_i m_{\nu_i}^{4}\text{log}\frac{m_{\nu_i}^{2}}{\mu^{2}}+\theta (\mu - M_{R})  M_{N}^{4}\text{log}\frac{M_{N}^{2}}{\mu^{2}}\right),
\end{align}
The matching of the complete and effective theory at threshold requires one to introduce a threshold contribution below $M_R$, 
$\Delta_{\text{TH}}V=-\frac{1}{32\pi^{2}}\left(M_{N}^{4}\text{log}\frac{M_{N}^{2}}{M_{R}^{2}}\right)$, whose expansion gives the threshold corrections to the 
$\mu^{2}$ and $\lambda$ parameters as $\Delta_{\text{TH}}\mu^{2}=\frac{1}{16\pi^{2}}|Y_{\nu}|^2 M_{R}^{2}$,
$\Delta_{\text{TH}}\lambda=-\frac{5}{32\pi^{2}}|Y_{\nu}|^{4}$.
Hence, we need to consider this shift in $\lambda$ at $\mu=M_{R}$ when solving the RGEs as
\begin{align}
 \lambda(M_{R})\to \lambda(M_{R})-\frac{5}{32\pi^{2}}|Y_{\nu}|^{4}
 \label{matching}
\end{align}
%Note that, for the (3,1) seesaw scheme, $n=1$ and $\text{Tr}(Y_\nu^{\dagger}Y_\nu)$ %should be replaced as $|Y_{\nu}|^2$.

\begin{figure}[h]
\centering
\includegraphics[width=0.75\textwidth]{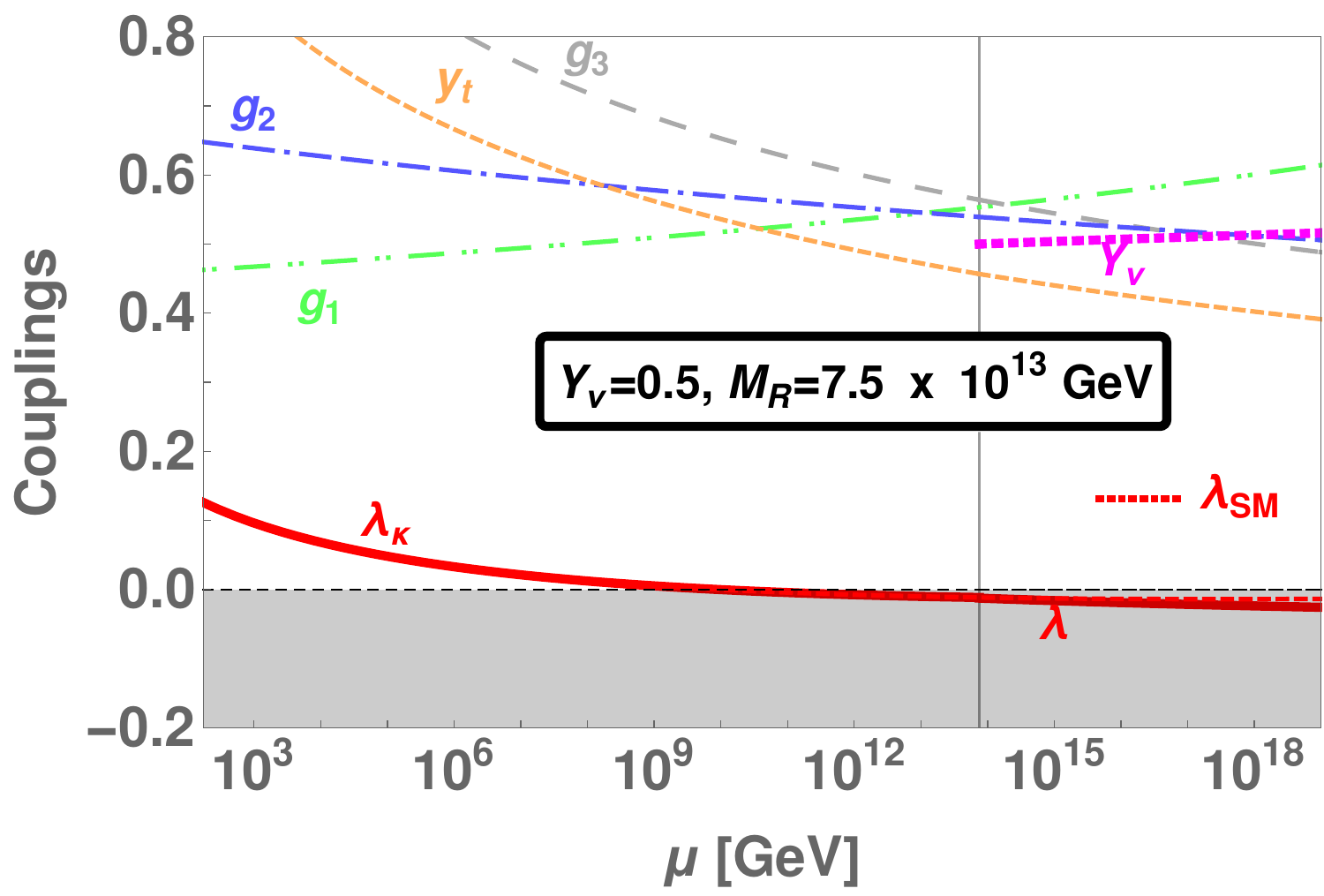}
\caption{\footnotesize{The continous (red) curve gives the evolution of the Higgs quartic self-coupling within the minimal (3,1) Type I seesaw scheme.
The gauge and Yukawa couplings $g_{1}$, $g_{2}$, $g_{3}$, $y_{t}$ and $Y_{\nu}$ are also indicated by the dashed lines.
The light neutrino mass is fixed at $m_\nu = 0.1$ eV, corresponding to a heavy neutrino mass $M_R$ of $7.5 \times 10^{13}$ GeV.
For comparison we show the evolution of the \sm coupling $\lambda_{\rm{SM}}$, seen as the red dashed line.
Finally, $\lambda_\kappa$ denotes the Higgs quartic coupling in the effective theory including neutrino mass through the Weinberg operator,
  while $\lambda$ is the corresponding quartic in the minimal missing-partner Type I seesaw theory.
\\
%
% \jv{\sout{The red dashed line denotes the \sm $\lambda$ running.}} \cy{\sout{$\lambda_\kappa$ is the Higgs quartic coupling in the effective seesaw theory with additional dimension five Weinberg operator. $\lambda_\kappa$ running almost coincides with $\lambda_{\text{SM}}$ running as Yukawa coupling $Y_\nu$ is not running below the scale $\mu=M_R$ and as the effect of $\kappa$ running is negligible. At the scale $\mu=M_R$, there is a negative shift which is coming from the Eq.~\ref{matching}, which is not visible because of relatively small Yukawa coupling. Above the scale $\mu=M_R$, $\lambda$ goes more negative than $\lambda_{\text{SM}}$ as Yukawa coupling starts to contribute, but the deviation is small. }}
}
} 
\label{RG-31-Running1}
\end{figure}

\begin{figure}[h]
\centering
\includegraphics[width=0.75\textwidth]{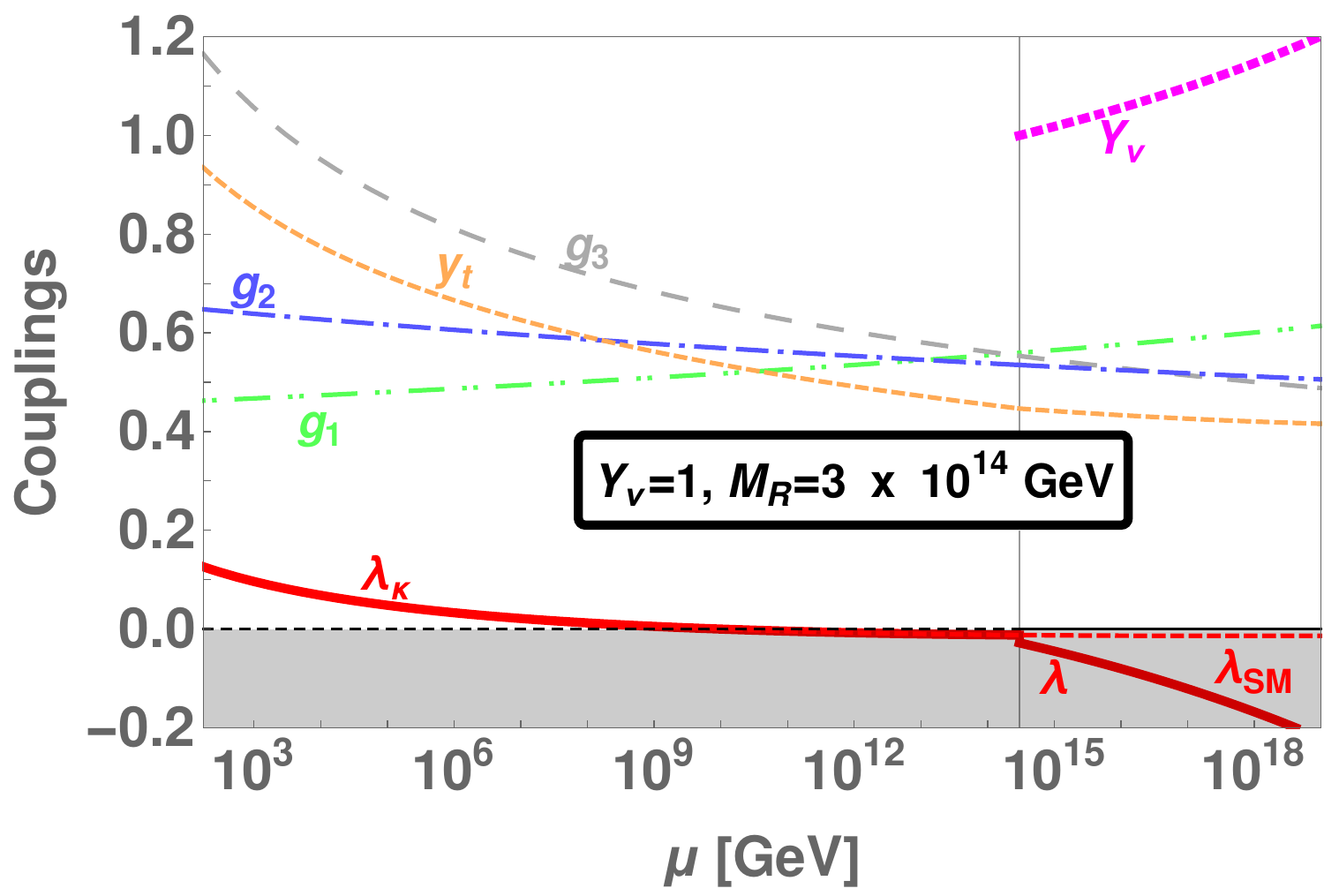}
\caption{\footnotesize{Same as Fig.~\ref{RG-31-Running1} but now with $Y_\nu=1$ and $M_R=3\times 10^{14}$.
    In contrast to  Fig.~\ref{RG-31-Running1}, $\lambda$ shows a marked deviation from $\lambda_{\rm{SM}}$ as a result of the larger Yukawa value, $Y_\nu = 1$.
    As before, $\lambda_{\rm{SM}}$ and $\lambda_\kappa$ nearly coincide as the Weinberg operator has negligible impact on RGE of the quartic Higgs coupling.  \\
    % \cy{\sout{The $\lambda_\kappa$ running is again the same as $\lambda_{\text{SM}}$ because of the same reason as previously mentoined. The negative shift at $\mu=3\times 10^{14}$~GeV is sizeable for large Yukawa coupling~(Eq.~\ref{matching}) and clearly visible in this case. Above this scale, the $\lambda$ goes negative shraplys compare to $\lambda_{\text{SM}}$ due to the effect of large Yukawa coupling and is clearly visible.}.}
  }}
\label{RG-31-Running2}
\end{figure}

Having set up our basic scheme, let us start by looking at the impact of the right-handed neutrinos on the stability of the Higgs vacuum. 
As we discussed at length in Section \ref{sec:sm-vac}, the \sm RGEs running of the Higgs quartic scalar coupling $\lambda_{\rm{SM}}$ is dominated by the top Yukawa,
which is the largest coupling present in the theory. As we saw, in this case the \sm $\lambda_{\rm{SM}}$ coupling becomes negative around the scale $\mu\sim 10^{10}$ GeV.
 However, within the seesaw completion, above the scale $\mu=M_{R}$ the neutrino Yukawa couplings $Y_\nu$ of
\eqref{seesaw-lag} can completely dominate the RGEs behavior of $\lambda$ as shown in Figs.~\ref{RG-31-Running1} and \ref{RG-31-Running2}. 
%
%%%%%%%%%%%%%%%%%%%%%%%%%%%%%%
%
%%%%%%%%%%%%%%%%%%%%%%%%%%%%%%%%%%5
Figs.~\ref{RG-31-Running1} and~\ref{RG-31-Running2} illustrates the effect of the new neutrino Yukawa coupling $Y_{\nu}$ on various other couplings. 
For illustration we have taken two representative values of Yukawa couplings $Y_{\nu}=0.5$ and 1. 
One sees how the problem of Higgs vacuum stability becomes more acute in a type-I seesaw completion of the Standard Model.
This was expected, since the addition of new fermions tends to destabilize the Higgs vacuum.
Notice that,  in the regime below the onset of the seesaw mechanism,  $\mu\leq M_R$, the running of the Higgs quartic coupling $\lambda_\kappa$ nearly coincides with $\lambda_{\rm{SM}}$.
  This follows due to the negligible effect of the Weinberg operator on the running of the Higgs quartic coupling.
The small negative shift in the $\lambda$ running at the scale $\mu=M_R$ results from the matching condition, which becomes clearly visible for larger Yukawa couplings, $Y_{\nu} \sim 1$.
Notice that in Figs.~\ref{RG-31-Running1} and \ref{RG-31-Running2} we have chosen a larger seesaw scale, with correspondingly larger ``Dirac'' neutrino Yukawa coupling values, in order to make the running coupling effects visible in the plots.

For $M_{N}\leq 10^{10}$~GeV, the Yukawa coupling is $Y_{\nu}\leq 10^{-3}$, hence too small to alter the running of $\lambda$ significantly.
As a result, the vacuum instability problem will persist. However, if regarded as an effective one, the theory remains mathematically self-consistent.
For larger $M_{N}$, for example $M_N$ close to the unification scale, the type-I seesaw relation~\ref{seesaw} implies that $Y_{\nu}$ should  also be sizeable. 
Such large Yukawa coupling will have a destabilizing effect, worsening the metastability of the \sm vacuum~\footnote{For example, if $M_{N} \sim 10^{14}$~GeV~(which implies $Y_{\nu} \sim 1$ for $m_{\nu} \sim 0.1$~eV), the vacuum lifetime is less than the age of the universe $\tau_U$, hence \sm metastability is worsened by the effect of this large Yukawa coupling.}. In fact, now the vacuum can be completely unstable, making the model inconsistent.
%
% \begin{figure}[h]
% \centering
% \includegraphics[width=0.75\textwidth]{HighScaleConsistency.pdf}
% \caption{\footnotesize{Zoomed view of the evolution of the Yukawa coupling $Y_{\nu}$ and the quartic Higgs self-coupling $\lambda$ within the minimal (3,1) seesaw. The Yukawa $Y_{\nu}$ is chosen small enough so that the seesaw provides at least a self-consistent effective theory. \cy{In this case \sm $\lambda$ running coincides with red solid line.}}} 
% \label{RG-31a-Running}
% \end{figure}

In conclusion, in seesaw scenarios the stability properties of the electroweak vacuum will at best be those of the \sm Higgs vacuum. 
In order to enhance Higgs vacuum stability it is desirable to further extend or embed the type-I seesaw~\cite{Ghosh:2017fmr}.
A natural way to do this is to assume spontaneous breaking of lepton number, as we do next.

%%%%%%%%%%%%%%%%%%%%%%%%%%%%%%%%%%%%%%%%%%%%%%%%%%%%
% \section{type-I seesaw with majoron}
% \label{sec:maj}

\section{The majoron completion }
\label{sec:maj-intro}

%%%%%%%%%%%%%%%%%%%%%%%%%%%%%%%%%%%%%%%%%%%%%%%%%%%%

We now consider the type-I seesaw extensions of the Standard Model, in which lepton number is promoted to a spontaneously broken symmetry within the 
\SM gauge framework~\cite{Chikashige:1980ui,Schechter:1981cv}. In addition to the right-handed neutrinos $\nu_{R}$ we add 
a complex scalar singlet $\sigma$ carrying two units of lepton number. The relevant Lagrangian is given by
\begin{align}
  \label{eq:maj-seesaw}
 \mathcal{L}=-\sum_{a,i} Y_{\nu}^{ai}\bar{\ell}_{L}^{a}\tilde{H}\nu_{R_i} - \frac{1}{2}\sum_{i,j} Y^{ij}_{R}\sigma\overline{\nu_{R_i}^{c}}\nu_{R_j}+\text{H.c.}
\end{align}
The resulting neutrino mass matrices in $\nu_{L}$ and $\nu_{R}$ basis is given by 
\begin{align}
  \mathcal{M}_{\nu}=
 \begin{pmatrix}
 0 & \frac{Y_{\nu}v_{H}}{\sqrt{2}} \\
 \frac{Y_{\nu}^{T}v_{H}}{\sqrt{2}} & \frac{Y_{R}v_{\sigma}}{\sqrt{2}} \\
 \end{pmatrix}
\end{align}
The effective light neutrino mass obtained by perturbative diagonalization of the above mass matrix is of the form
\begin{align}
 m_{\nu}\simeq Y_{\nu}Y_{R}^{-1}Y_{\nu}^{T}\frac{v_{H}^{2}}{\sqrt{2}v_{\sigma}}
 \label{seesaw formula in type-I seesaw with majoron}
\end{align}

In the presence of the complex scalar singlet $\sigma$, the most general Higgs potential that can drive electroweak and lepton number symmetry breaking 
is given by~\cite{Joshipura:1992hp}
\begin{align}
  V(\sigma,H)=-\mu_{H}^{2}H^{\dagger}H-\mu_{\sigma}^{2}\sigma^{\dagger}\sigma+\lambda_H (H^{\dagger}H)^{2}+\lambda_{\sigma}(\sigma^{\dagger}\sigma)^{2}
  +\lambda_{H\sigma}(H^{\dagger}H)(\sigma^{\dagger}\sigma).
  \label{eq:maj-pot}
\end{align}

This potential is bounded from below if $\lambda_{\sigma}$, $\lambda_H$ and $\lambda_{H\sigma}+2\sqrt{\lambda_{\sigma}\lambda}$ are all positive. 
In addition to the standard \SM gauge invariance, in the unbroken phase, the theory is also invariant under lepton number. 
The above potential can develop a minimum for non-zero vacuum expectation values of both H and $\sigma$ if 
$\lambda_H$, $\lambda_{\sigma}$ and $4\lambda_H\lambda_{\sigma}-\lambda_{H\sigma}^{2}$ are all positive. 
The vevs break both the electroweak and lepton number symmetries, three of the degrees of freedom are eaten by the massive Standard Model gauge bosons, 
while the imaginary part of the $\sigma$ corresponds to the physical majoron $J=Im~\sigma$. 
The real parts of $H$ and $\sigma$ will mix with each other to give two CP-even mass eigenstates $h_{1}$ and $h_{2}$. 
The lighter of these is identified with 125 GeV Higgs boson~\cite{Aad:2012tfa,Chatrchyan:2012xdj}. 

\section{Vacuum stability in type-I seesaw with majoron}
\label{sec:result-maj}

%%%%%%%%%%%%%%%%%%%%%%%%%%%%%%%%%%%%%%%%%%%%%%%%%%%%

Here we take again the simplest majoron extension of the type-I seesaw mechanism based on the (3,1) missing partner scheme considered above
\footnote{Note that vacuum stability in a seesaw majoron model was discussed in~\cite{Sirkka:1994np}.
    However, the majoron in that paper was completely detached from the neutrino sector, lacking any solid motivation.
    Moreover, the low scale choice for $v_\sigma$ was artificial, requiring tiny Dirac Yukawa couplings.
    In our opinion, it is best to present the discussion within a genuine low-scale neutrino mass generation mechanism, as in Ref.~\cite{Bonilla:2015kna}}.
We adopt the high-scale seesaw limit $v_{\sigma} \gg v_{H}$.
In this limit, mass of the heavier CP-even scalar boson and right neutrino are 
approximately given as $m_{h_2} \equiv M \approx \sqrt{2\lambda_{\sigma}v_{\sigma}}$ and $M_{N}\approx \frac{Y_{R}}{\sqrt{2}}v_{\sigma}$.
The light and heavy Higgs sectors will be almost decoupled, though we can still allow appreciable $\lambda_{H\sigma}$ with 
very small mixing angle $\alpha$, see Appendix~\ref{app:higgs}. 
For simplicity, we consider nearly degenerate $M$ and $M_{N}$, such that we have only one threshold scale
$\mu=M_{N}\,\text{or}\,M$, below which the theory is an effective one.
Above that scale we have the full theory with all the new couplings running. 
\begin{figure}[h]
\centering
\includegraphics[width=0.25\textwidth]{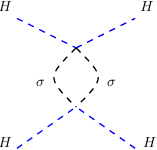}~~~~~~~~~~
\includegraphics[width=0.25\textwidth]{LoopLambda.pdf}
\caption{\footnotesize{One-loop correction to the Higgs quartic interaction in type-I seesaw with majoron.}} 
\label{One loop corrections to lambda in majoron case}
\end{figure}

When going from energies above $M$ to energies below $M$ we need to integrate out the massive scalar degree of freedom at tree level as described in Ref.~\cite{EliasMiro:2012ay}. 
This leads to a tree level threshold effect that arises from the matching conditions at the energy scale $\mu=M$. 
We will now briefly describe this procedure. We can write the scalar potential for the case of type-I seesaw with majoron extensions as,
\begin{align}
 V_{0}=\lambda_H \left(H^{\dagger}H-\frac{v_H^2}{2}\right)^2+\lambda_\sigma \left(\sigma^{\dagger}\sigma-\frac{v_\sigma^2}{2}\right)^2+\lambda_{H\sigma}\left(H^{\dagger}H-\frac{v_H^2}{2}\right)
 \left(\sigma^{\dagger}\sigma-\frac{v_{\sigma}^2}{2}\right)
 \label{minimum potential majoron}
\end{align}

In our case, $v_{\sigma} \gg v_{H}$, therefore $M$ is much larger than the Higgs mass $m_{h_1}$. As a result,
below the scale $\mu=M$, we can integrate out the field $\sigma$ using the following equation of motion~(apart from derivative terms)
\begin{align}
 & 2\lambda_\sigma\left(\sigma^{\dagger}\sigma-\frac{v_\sigma^2}{2}\right)+\lambda_{H\sigma}\left(H^{\dagger}H-\frac{v_H^2}{2}\right)=0\nonumber\\
 & \sigma^{\dagger}\sigma=\frac{v_\sigma^2}{2}-\frac{\lambda_{H\sigma}}{2\lambda_\sigma}\left(H^{\dagger}H-\frac{v_H^2}{2}\right)
 \label{field integrate out}
\end{align}

In order to obtain the effective potential below the scale $\mu=M$, we use Eq.~\ref{field integrate out} in Eq.~\ref{minimum potential majoron},
leading to the effective Higgs potential expressed as
\begin{align}
 V_{\text{eff}}=\lambda_H^{'}\left(H^{\dagger}H-\frac{v_{H}^{2}}{2}\right)^{2},
\end{align}
where $\lambda_H^{'}$ is identified as

\begin{align}
\lambda_H^{'}\equiv\lambda_{\kappa}=\lambda_H-\frac{\lambda_{H\sigma}^{2}}{4\lambda_{\sigma}}.
\label{shift}
\end{align}
Notice that, since only the dimension-five Weinberg operator is running below the scale $\mu=M$, one has that the running of $\lambda_H^{'}$ is essentially the same as that of
  $\lambda_\kappa$ in the effective type-I seesaw.
  Moreover, at tree-level the numerical value of $\lambda_{H}^{'}(M_Z)$ and $\lambda_{\text{SM}}(M_Z)$ is the same, since in both cases one must reproduce the 125 GeV Higgs mass.
Eq.~\ref{shift} suggests that the matching condition at the scale $\mu=M$ induces a shift in the Higgs quartic coupling, $\delta\lambda=\frac{\lambda_{H\sigma}^{2}}{4\lambda_{\sigma}}$.
This corresponds to a larger Higgs quartic coupling above the scale $M$ and improves the chances of keeping $\lambda_H$ positive all the way.

In Appendix~\ref{app:maj}, we give the two-loop RGEs of the full theory.
Note that below the scale $M$ the RGEs are the ones with $\beta_{\lambda_{H\sigma}}$ and $\beta_{\lambda_{\sigma}}$ removed and $\lambda_H$ replaced by $\lambda_H'$.
Above $M$ one needs to include $\beta_{\lambda_{H\sigma}}$, $\beta_{\lambda_{\sigma}}$ and find $\lambda_H$ using the full RGEs with the boundary condition as in Eq.~\ref{shift} at $\mu=M$.
As far as the new Yukawa couplings are concerned, they can be obtained by substituting $Y_{\nu}=\theta\left(\mu-M_{N}\right)Y_{\nu}$ and $Y_{R}=\theta\left(\mu-M\right)Y_{R}$
on the right side of the RGEs of the full theory.
\begin{figure}[h!]
\centering
\includegraphics[width=0.49\textwidth]{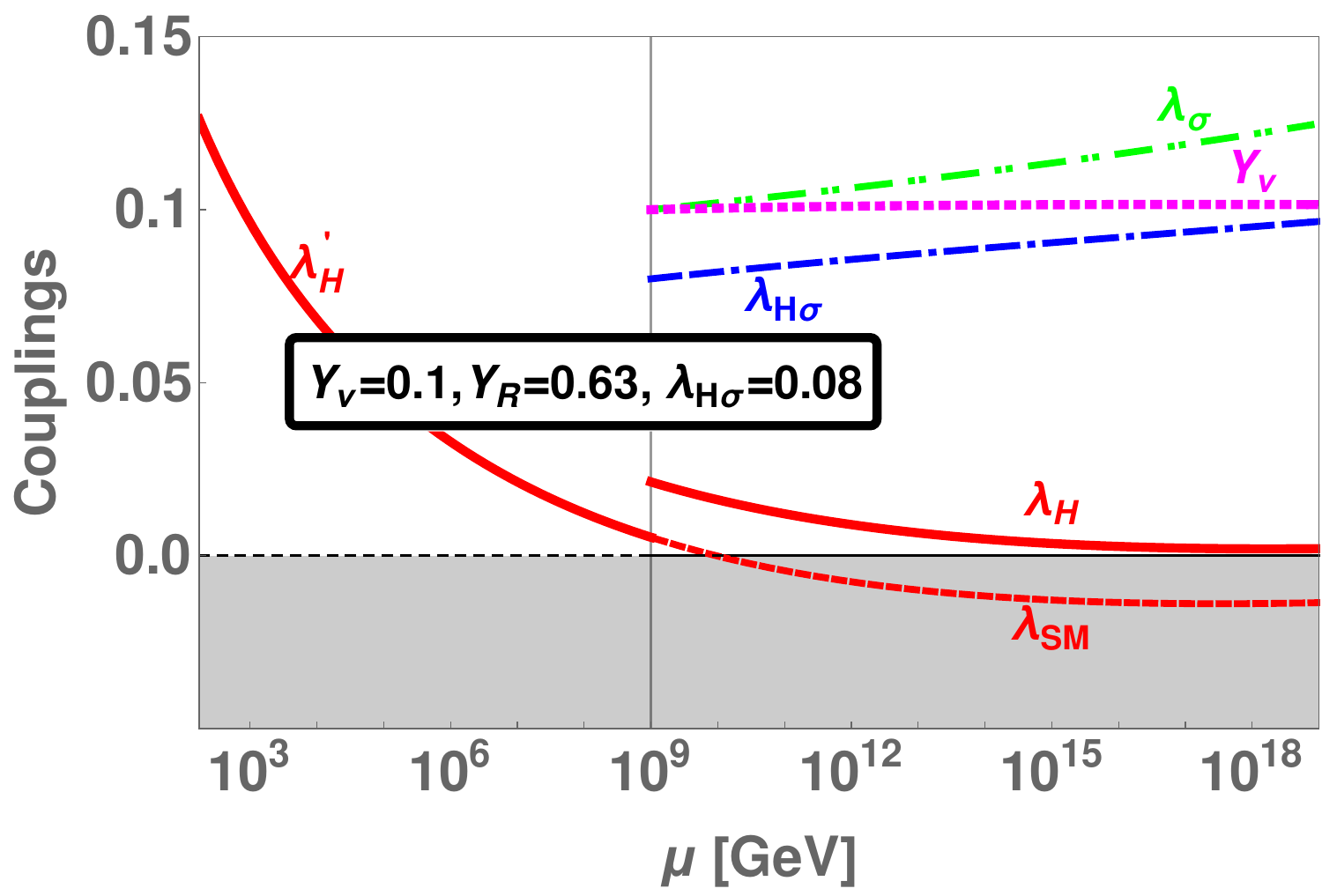}
\includegraphics[width=0.49\textwidth]{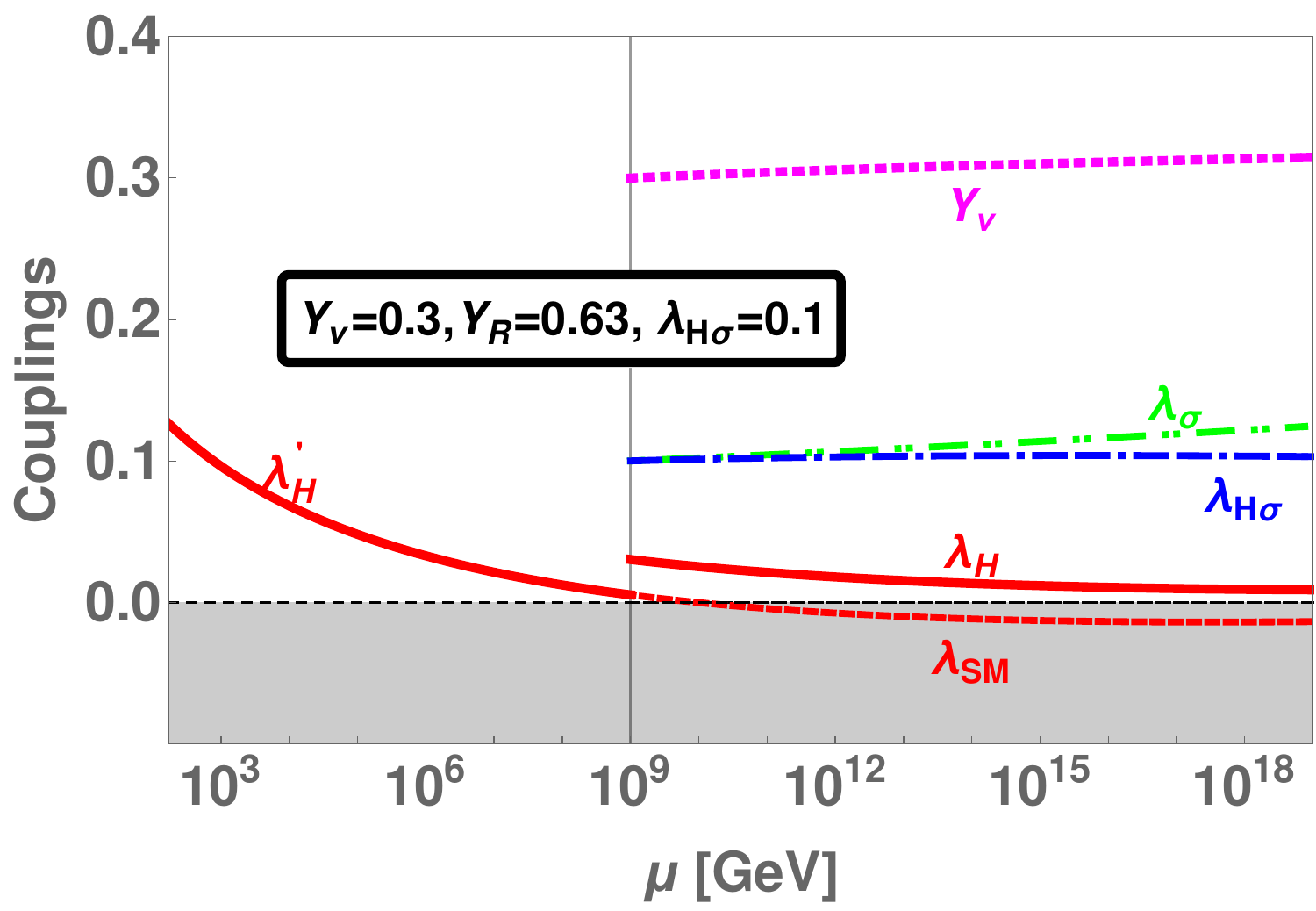}
\includegraphics[width=0.49\textwidth]{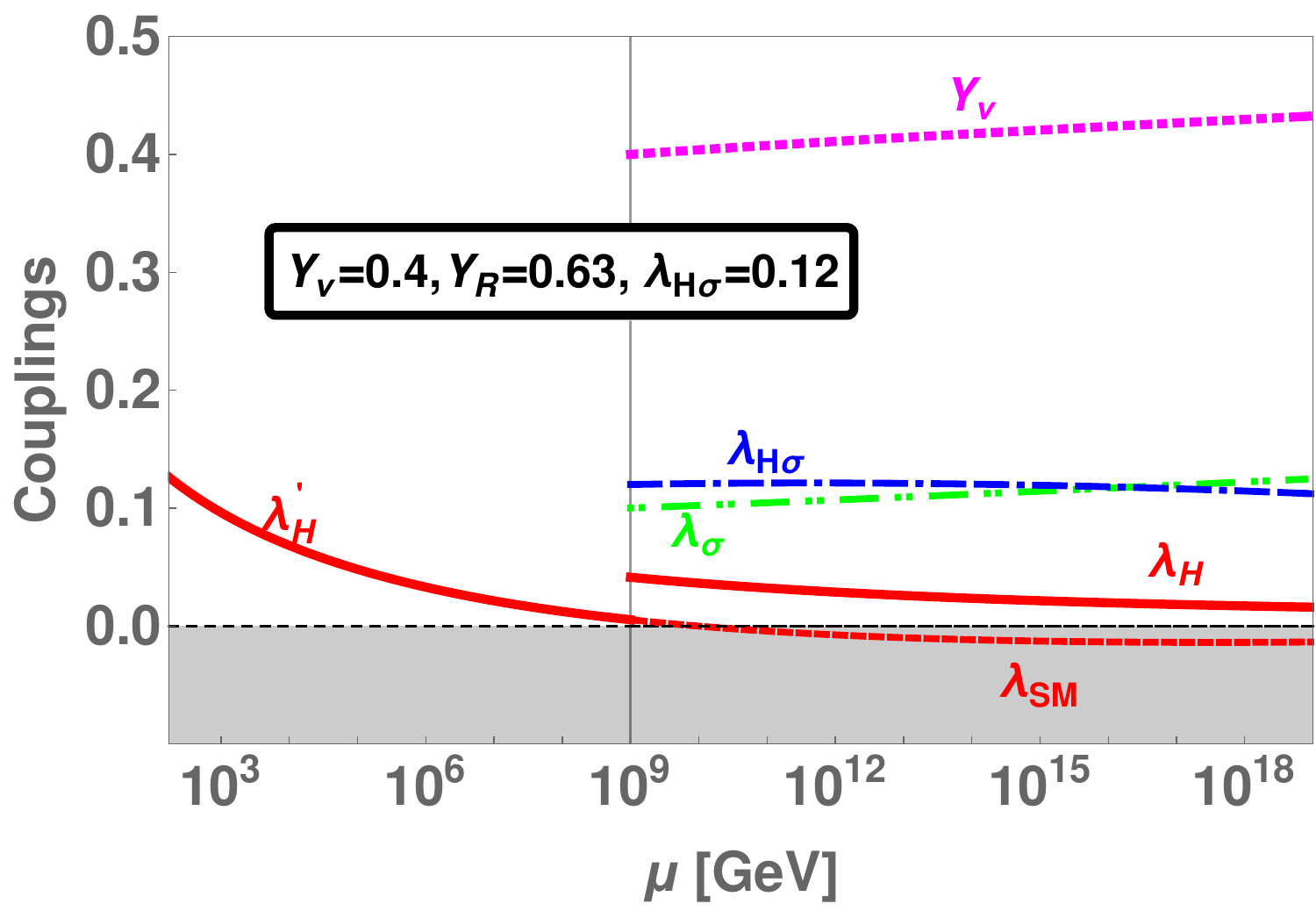}
\includegraphics[width=0.49\textwidth]{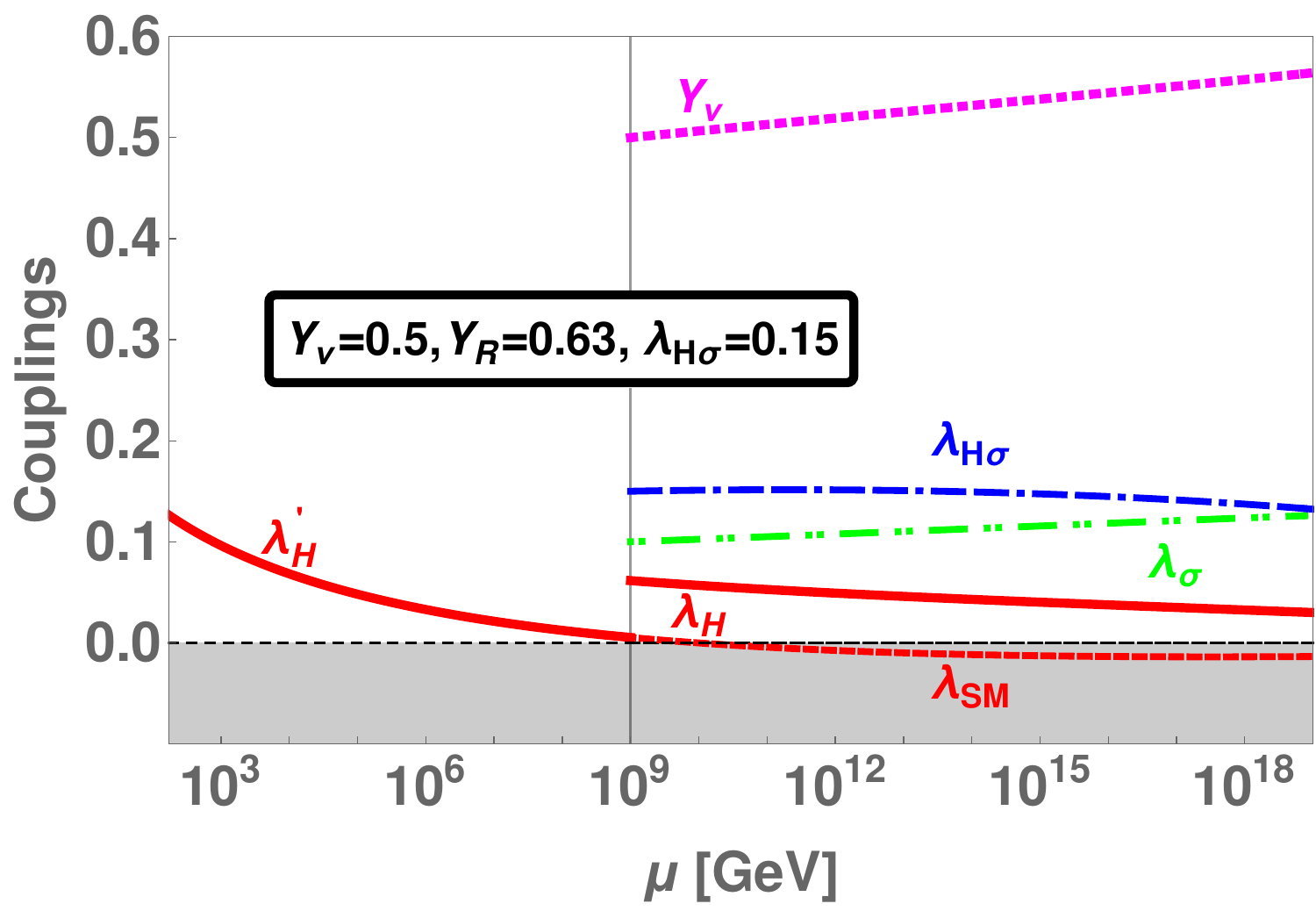}
\caption{\footnotesize{Evolution of the scalar quartic couplings $\lambda_H$, $\lambda_{\sigma}$ and $\lambda_{H\sigma}$ in type-I seesaw with majoron.
      We include in purple dashed color the Dirac Yukawa coupling $Y_{\nu}$, see Eqs.~\eqref{eq:maj-seesaw} and \eqref{eq:maj-pot}. 
      For comparison with \sm we have also shown the RGE for $\lambda_{\rm{SM}}$, red dashed curve.
      Here $\lambda'_H$ is the effective Higgs quartic coupling below the mass threshold of the heavy particles and is essentially the same as $\lambda_{\kappa}$ in the effective type-I seesaw,
      see Eq.~\ref{shift}. 
Since, in this regime the RGE of $\lambda'_H$ differs from that of $\lambda_{\rm{SM}}$ only due to the tiny contribution of the effective Weinberg operator, $\lambda'_H$ and 
$\lambda_{\rm{SM}}$ almost coincide with each other. 
See text for more detailed discussion of various key features of the plots. 
\\
%\jv{\sout{The red dashed lines correspond to the \sm running of the quartic  $\lambda_{\text{SM}}$ coupling}}. \cy{\sout{$\lambda_{H}^{'}$ is the quartic Higgs coupling in effective Majoron seesaw theory %and its running almost coincides with the $\lambda_{\text{SM}}$ running as all the new couplings are not running below the scale $\mu=M$. At the scale $\mu=M$, there is a jump in $\lambda_H$ due to the %threshold coming from Eq.~\ref{shift}. Combine effects of this jump and running of $\lambda_{H\sigma}$ are keeping the $\lambda_H$ positive all the way up to Planck scale inspite of large Yukawa %coupling.}}
}} 
\label{RG-31aJ-Running}.
\end{figure}

Our results for the (3,1) type-I seesaw mechanism with majoron are shown in Fig.~\ref{RG-31aJ-Running}, where we have taken $M\approx M_{N}\approx 10^{9}$~GeV, such that the threshold effects 
start contributing positively to $\lambda_H$ before the \sm instability scale $\Lambda_{SM} \approx 10^{10}$~GeV. 
We have taken $\lambda_{\sigma}=0.1$ at the scale $M$.
The renormalization group evolution in Fig.~\ref{RG-31aJ-Running} is shown for four Yukawa coupling values $Y_{\nu}=0.1,\,0.3,\,0.4\,\text{and}\,0.5$. 
It shows that, indeed, the stability properties can substantially improve due to the presence of the new scalar.
In fact, for appreciable Yukawa couplings, one can have positive $\lambda_H$ all the way up to Planck scale. 
For the Yukawa couplings $Y_{\nu}=0.1,\,0.3,\,0.4\,\text{and}\,0.5$, the required values of minimum $\lambda_{H\sigma}$ are 0.08, 0.1, 0.12 and 0.15 respectively.

%%%%%%%%%%%%%%%%%%%%%%%%%%%%%%%%%%%%%%%%%%%%%%%%%%%%
\section{Comparing standard and missing partner type-I seesaw}
\label{sec:3gen}
%%%%%%%%%%%%%%%%%%%%%%%%%%%%%%%%

So far we have taken the missing partner seesaw mechanism based on the (3,1) construction as our benchmark.
This choice was made for the reasons given at the end of Sec.~\ref{sec:miss-seesaw}. 
Such scheme can be made phenomenologically viable in the presence of radiative corrections associated, for example, to a dark matter completion~\footnote{
An explicit scotogenic model of this type has been proposed in~\cite{Rojas:2018wym}.}.

Here we compare the stability properties of this simplest benchmark with those of a missing partner seesaw based on (3,2) construction and with those of the standard (3,3) type-I seesaw mechanism.
For completeness we also compare with the \sm stability results.

\begin{figure}[h]
\centering
\includegraphics[width=0.7\textwidth]{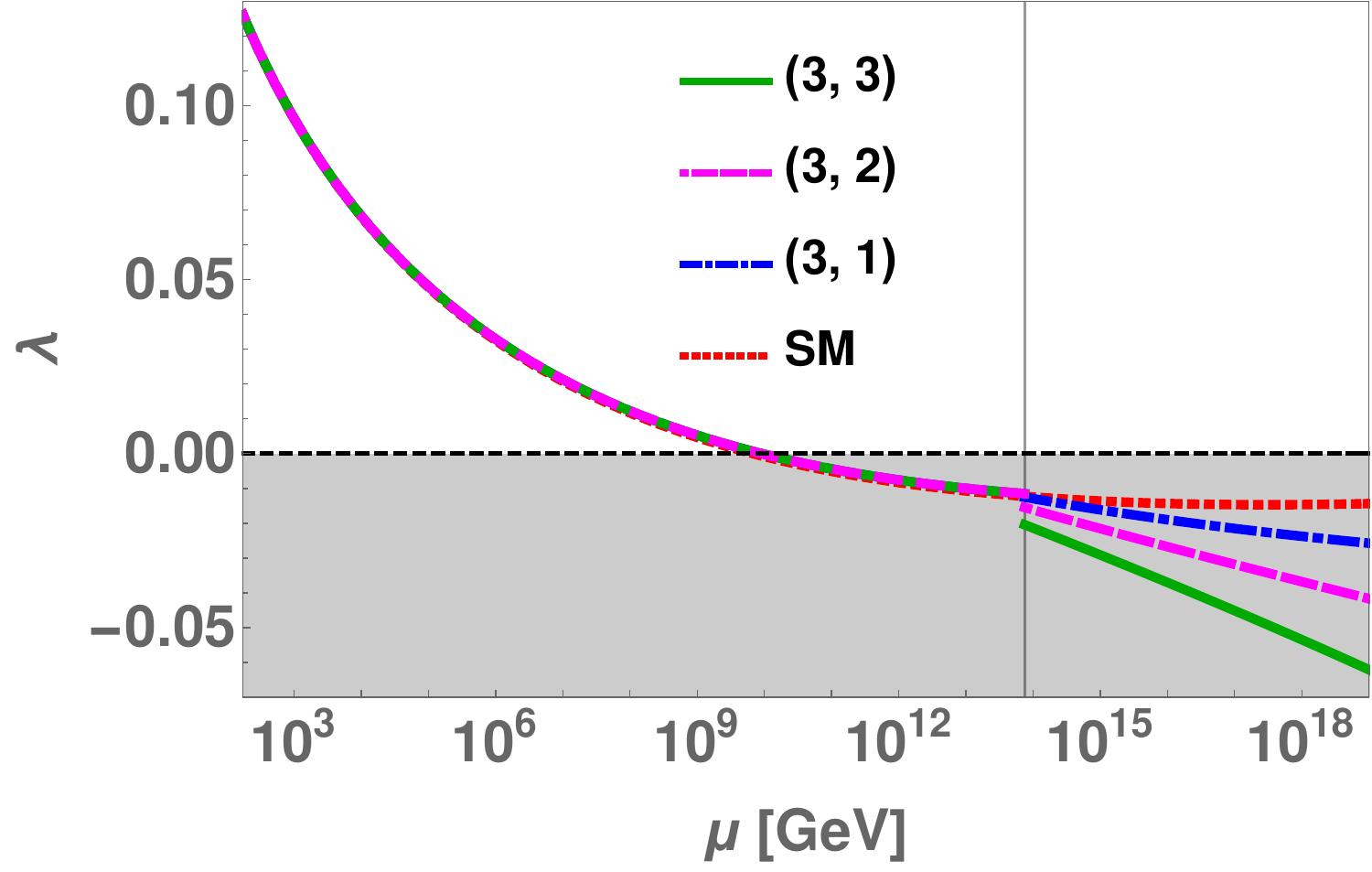}
\caption{\footnotesize{Zoomed view of the evolution of the quartic Higgs self-coupling $\lambda$ in the \sm (red-dashed) and in the (3,1), (3,2) and (3,3) seesaw extensions (blue dot-dashed, magenta-dashed and green solid, respectively). In the (3,1) case we have taken $Y_\nu = 0.5$, while for $(3,n=2,3)$ we took $Y^{aj}_\nu = 0.5$; $a=j= 1,\,..,n$ and $Y^{aj}_\nu = 0$ for $a \neq j$.}}
\label{robustness}
\end{figure}

\begin{figure}[h]
\centering
\includegraphics[width=0.7\textwidth]{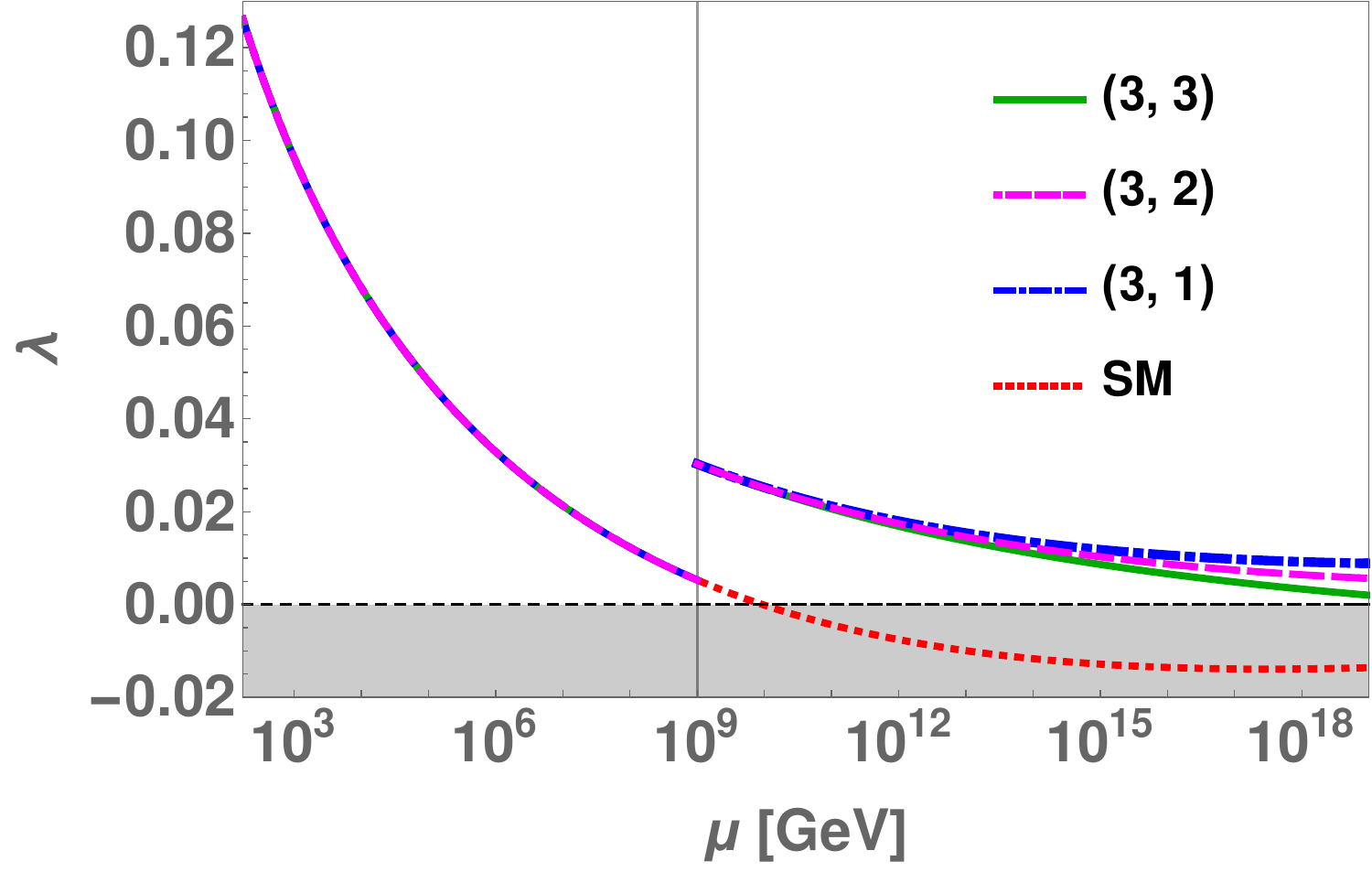}
\caption{\footnotesize{Zoomed view of the evolution of the quartic Higgs self-coupling $\lambda$ in the \sm (red-dashed) and (3,1), (3,2) and (3,3) majoron seesaw (blue dot-dashed, magenta-dashed and green solid, respectively). In the (3,1) case we have taken $Y_\nu = 0.3$, while for $(3,n=2,3)$ we took $Y^{aj}_\nu = 0.3$; $a=j= 1,\,..,n$ and $Y^{aj}_\nu = 0$ for $a \neq j$.}}
\label{fig:maj-robustness}
\end{figure}

As already mentioned, the problem of Higgs vacuum stability in type-I seesaw extensions only gets worse with the addition of extra right-handed neutrinos. 
This fact is clearly illustrated in Fig. \ref{robustness}, where we compare the evolution of the Higgs quartic self-coupling $\lambda$ within the \sm within the (3,$n$) seesaw completions, with 
$n$ = 1, $n$ = 2, and $n$ = 3. 
Note that, for the general $(3,n)$ seesaw scheme $|Y_{\nu}|^2$ should be replaced as $\text{Tr}(Y_\nu^{\dagger}Y_\nu)$ and Eq.~\ref{matching} should be replaced as 
  \begin{align}
 \lambda(M_R)\to \lambda(M_R)-\frac{5 n^2}{32\pi^2}\text{Tr}(Y_\nu^{\dagger}Y_\nu)^2,
\end{align}
where $n$ is the number of right handed neutrinos.
For simplicity, in Fig. \ref{robustness}, we have fixed the benchmark value of $Y^{aj}_\nu = 0.5$; $a=j= 1,2,3$ and taken the off-diagonal terms to be zero for the $(3,3)$ case. Note that such a choice is unrealistic vis-a-vis neutrino oscillation data. However, taking the Yukawa texture consistent with neutrino oscillation data will not change our conclusions. Therefore, for the sake of simplicity, we have taken this simple choice. 
\\
In contrast, going to the  (3,3) majoron type-I seesaw with three right handed neutrinos, we find that the Higgs vacuum can be still kept stable up to Planck scale for appreciable Yukawa couplings. 
Of course, the presence of additional fermions means that the maximum values of $Y^{ai}_\nu$, for which Higgs vacuum stability can be achieved up to Planck scale, is somewhat reduced.
In Fig.~\ref{fig:maj-robustness} we compare the Higgs vacuum stability of the (3,3) majoron seesaw case with its (3,1) analogue as well as with Standard Model. 

For Fig.~\ref{fig:maj-robustness} we have taken $Y_\nu = 0.3$ for the (3,1) case, while for the (3,3) case we have taken $Y^{ai}_\nu = 0.3$; $a=i=1,2,3$, with all the off-diagonal entries taken to zero.
The remaining parameters are kept the same as described previously for the (3,1) majoron seesaw case.
In short, phenomenologically realistic type-I seesaw majoron models can have a stable vacuum all the way up to Planck scale. 
\section{Summary and outlook}
\label{sec:summary-discussion}

We have examined the consistency of electroweak symmetry breaking within the context of the simplest high-scale type-I seesaw mechanism. 
We have derived the full two-loop RGEs for the relevant parameters, such as the quartic Higgs self-coupling $\lambda$ of the Standard Model within the schemes of interest.
These are compared, for calibration, with the \sm results.
The addition of fermionic fields like  ``right-handed'' neutrinos, has a destabilizing effect on the Higgs boson vacuum.
For the simplest type-I seesaw with bare mass term for the right-handed neutrinos, we found that for sizeable Yukawa couplings the Higgs quartic self-coupling $\lambda$ becomes 
negative much before reaching the seesaw scale. For such ``large'' Yukawas the type-I seesaw may be inconsistent even as an effective theory. 
We have taken as our simplest benchmark neutrino model the ``incomplete'' (3,1) seesaw scheme with a single right-handed neutrino, as it has the ``best'' stability properties within the class of high-scale type-I seesaw schemes.
We compared this case, in which only one oscillation scale is generated at tree level, with the ``higher'' (3,2) type-I seesaw, in which the other mass scale also arises from the tree level seesaw mechanism.
In both ``missing partner'' type-I seesaw schemes,  (3,1) and (3,2), the \znbb prediction given in Fig.\ref{fig:dbd} holds.
We also studied the stability of the electroweak vacuum for the canonical sequential (3,3) type-I seesaw, in which all three neutrinos get tree-level mass.
We showed how the stability properties improve in the case of spontaneous lepton number violation due to the presence of a Nambu-Goldstone boson, the majoron. 

To sum up, our results show how, in contrast to the type-I seesaw with explicit breaking of lepton number, the majoron version can have stable electroweak vacuum all 
the way upto Planck scale for reasonable Yukawa coupling choices. Thus, the majoron completion of type-I seesaw schemes can be considered as fully consistent theories. 

Before concluding we should note the cosmological advantages of the majoron completion.
The first is that it can also provide a dark matter candidate, namely the majoron~\cite{Berezinsky:1993fm}, providing an alternative to the $\Lambda$CDM paradigm. 
The majoron is assumed to get mass from gravitational effects that explicitly violate the global lepton number~\cite{coleman:1988tj}.
Assuming that its mass lies in the keV range one can show that it can provide a viable warm dark matter candidate.
It decays to neutrinos, with a tiny strength proportional to their mass~\cite{Schechter:1981cv}.
Hence, it is naturally long-lived on a cosmological scale, as required, with lifetime $\tau_J$ larger than the age of the Universe $t_0 =13.8\,\mathrm{Gyr}\simeq 4\times10^{17}\,\mathrm{s}$.
Such majoron dark matter scenario has been shown to be consistent with cosmic microwave background data for adequate choices of the relevant parameters~\cite{Lattanzi:2007ux,Lattanzi:2013uza,Audren:2014bca}, the majoron decay lifetime constraints ranging from $\tau_J > 50-160\,\mathrm{Gyr}$.
Using N-body simulations one can also show that majoron dark matter provides a viable alternative to the $\Lambda$CDM scenario, with predictions that can differ substantially on small scales~\cite{Kuo:2018fgw}.
Finally we also mention that, in addition to dark matter, the majoron picture may also provide new insights to other cosmological challenges of the Standard Model, such as inflation~\cite{Boucenna:2014uma} and leptogenesis~\cite{Sierra:2014sta}.

%%%%%%%%%%%%%%%%%%%%%%%%%%%%%%%%%%%%%%%%%%%%%%%%%%%%%

\begin{acknowledgments}

  This work is supported by the Spanish grant FPA2017-85216-P (AEI/FEDER, UE), PROMETEO/2018/165 (Generalitat Valenciana) and the Spanish Red Consolider MultiDark FPA2017-90566-REDC.
  We thank Martin Hirsch and Werner Porod for useful discussions.

\end{acknowledgments}

%%%%%%%%%%%%%%%%%%%%%%%%%%%%%%%%%%%%%%%%%%%%%%%%%
\appendix
\section{Higgs Sector in majoron Model}
\label{app:higgs}
%%%%%%%%%%%%%%%%%%%%%%%%%%%%%%%%%%%%%%%%%%%%%%%%%

The scalar potential for the majoron type-I seesaw is given by,
\begin{align}
 V=-\mu_{H}^{2}H^{\dagger}H-\mu_{\sigma}^{2}\sigma^{\dagger}\sigma+\lambda_H (H^{\dagger}H)^{2}+\lambda_{\sigma}(\sigma^{\dagger}\sigma)^{2}+\lambda_{H\sigma}(H^{\dagger}H)(\sigma^{\dagger}\sigma).
 \label{scalar-maj}
\end{align}
The \sm gauge singlet scalar $\sigma$ carries two units of lepton number and its vev 
$\langle \sigma \rangle  = \frac{v_\sigma}{\sqrt{2}}$ breaks the lepton number symmetry $U(1)_L $ to a $\mathbb{Z}_2$ subgroup.
After symmetry breaking one has, in the unitary gauge 
\begin{align}
H\rightarrow \frac{1}{\sqrt{2}}
 \begin{pmatrix}
 0 \\
 v_{H}+h' \\
 \end{pmatrix}
,\hspace{1cm}
\sigma\to \frac{v_{\sigma}+\sigma'}{\sqrt{2}}.
\end{align}
The scalars $h'$ and $\sigma'$ will mix with each other, their mass eigenvalues are given by,
\begin{align}
m_{h_{1}}^{2}=\lambda_H v_{H}^{2}+\lambda_{\sigma} v_{\sigma}^{2}-\sqrt{(\lambda_H v_{H}^{2}-\lambda_{\sigma} v_{\sigma}^{2})^{2}+(\lambda_{H\sigma}v_{H} v_{\sigma})^{2}} \, , \\
m_{h_{2}}^{2}=\lambda_H v_{H}^{2}+\lambda_{\sigma} v_{\sigma}^{2}+\sqrt{(\lambda_H v_{H}^{2}-\lambda_{\sigma} v_{\sigma}^{2})^{2}+(\lambda_{H\sigma}v_{H} v_{\sigma})^{2}} \, .
\end{align}

The mass eigenstates $h_1, h_2$ are related to the fields $h', \sigma'$ by the mixing matrix parameterized by the angle $\alpha$ and given by
\begin{align}
 \begin{pmatrix}
  h_{1} \\
  h_{2}  \\
 \end{pmatrix}
=
\begin{pmatrix}
 \text{cos}~\alpha & -\text{sin}~\alpha \\
 \text{sin}~\alpha & \text{cos}~\alpha \\
\end{pmatrix}
\begin{pmatrix}
 h' \\
 \sigma' \\
\end{pmatrix} \, ,
\end{align}
where the mixing angle $\alpha$ is given by,
\begin{align}
\text{sin}~2\alpha=\frac{\lambda_{H\sigma}v_{H}v_{\sigma}}{\sqrt{(\lambda_H v_{H}^{2}-\lambda_{\sigma}v_{\sigma}^{2})^{2}+(\lambda_{H\sigma}v_{H}v_{\sigma})^{2}}} \, ,\nonumber  \\
\text{cos}~2\alpha=\frac{\lambda_H v_{H}^{2}-\lambda_{\sigma}v_{\sigma}^{2}}{\sqrt{(\lambda_H v_{H}^{2}-\lambda_{\sigma}v_{\sigma}^{2})^{2}+(\lambda_{H\sigma}v_{H}v_{\sigma})^{2}}} \, .
\label{eq:higgs-mix}
\end{align}

One can see from \eqref{eq:higgs-mix} that in the limit $v_\sigma \gg v_H $ the mixing angle $\alpha \to 0$, irrespective of the value of the quartic couplings.

%%%%%%%%%%%%%%%%%%%%%%%%%%%%%%%%%%%%%%%%%%%%%%%%%%%%%%%%%%%%%%%%%%%%%%%%
\section{RGEs: Type I seesaw}
\label{app:seesaw}
%%%%%%%%%%%%%%%%%%%%%%%%%%%%%%%%%%%%%%%%%%%%%%%%%%%%%%%%%%%%%%%%%%%%%%%%

We have used the package SARAH~\cite{Staub:2015kfa} to do the RGEs analysis in our work. The $\beta$ function of a given parameter c is given by
\begin{align*}
 \frac{dc}{dt}\equiv\beta_{c}=\frac{1}{16\pi^{2}}\beta_{c}^{(1)}+\frac{1}{(16\pi^{2})^{2}}\beta_{c}^{(2)} \, .
\end{align*}
where $\beta_{c}^{(1)}$ are the one-loop RGEs corrections and $\beta_{c}^{(2)}$ are the two-loop RGEs corrections.

%%%%%%%%%%%%%%%%%%%%%%%%%%%%%%%%%%%%%%%%%%%%%%%%%%%%%%%%%%%%%%%%%%%%%%%
\subsection{Higgs quartic scalar self coupling}
\label{app:seesaw-quartic}
%%%%%%%%%%%%%%%%%%%%%%%%%%%%%%%%%%%%%%%%%%%%%%%%%%%%%%%%%%%%%%%%%%%%%%%

For the most general (3,n) seesaw the one-loop and two-loop RGEs corrections to the Higgs quartic self-coupling are given by:
\begin{align} 
\beta_{\lambda}^{(1)} & =  
+\frac{27}{200} g_{1}^{4} +\frac{9}{20} g_{1}^{2} g_{2}^{2} +\frac{9}{8} g_{2}^{4} -\frac{9}{5} g_{1}^{2} \lambda -9 g_{2}^{2} \lambda +24 \lambda^{2} 
 +12 \lambda y_{t}^{2} +4 \lambda \mbox{Tr}\Big({Y_\nu  Y_{\nu}^{\dagger}}\Big) 
 -6 yt^{4} \nonumber \\
 &-2 \mbox{Tr}\Big({Y_\nu  Y_{\nu}^{\dagger}  Y_\nu  Y_{\nu}^{\dagger}}\Big) \, , 
 \end{align}
 \begin{align}
\beta_{\lambda}^{(2)} & =  
-\frac{3411}{2000} g_{1}^{6} -\frac{1677}{400} g_{1}^{4} g_{2}^{2} -\frac{289}{80} g_{1}^{2} g_{2}^{4} +\frac{305}{16} g_{2}^{6} 
+\frac{1887}{200} g_{1}^{4} \lambda +\frac{117}{20} g_{1}^{2} g_{2}^{2} \lambda -\frac{73}{8} g_{2}^{4} \lambda +\frac{108}{5} g_{1}^{2} \lambda^{2} \nonumber \\
&+108 g_{2}^{2} \lambda^{2} -312 \lambda^{3} -\frac{171}{100} g_{1}^{4} y_{t}^{2} 
 +\frac{63}{10} g_{1}^{2} g_{2}^{2} y_{t}^{2} -\frac{9}{4} g_{2}^{4} y_{t}^{2} 
 +\frac{17}{2} g_{1}^{2} \lambda y_{t}^{2} +\frac{45}{2} g_{2}^{2} \lambda y_{t}^{2} +80 g_{3}^{2} \lambda y_{t}^{2} \nonumber \\
 &-144 \lambda^{2} y_{t}^{2} 
 -\frac{9}{100} g_{1}^{4} \mbox{Tr}\Big({Y_\nu  Y_{\nu}^{\dagger}}\Big) -\frac{3}{10} g_{1}^{2} g_{2}^{2} \mbox{Tr}\Big({Y_\nu  Y_{\nu}^{\dagger}}\Big) 
 -\frac{3}{4} g_{2}^{4} \mbox{Tr}\Big({Y_\nu  Y_{\nu}^{\dagger}}\Big) +\frac{3}{2} g_{1}^{2} \lambda \mbox{Tr}\Big({Y_\nu  Y_{\nu}^{\dagger}}\Big) \nonumber \\
 &+\frac{15}{2} g_{2}^{2} \lambda \mbox{Tr}\Big({Y_\nu  Y_{\nu}^{\dagger}}\Big) -48 \lambda^{2} \mbox{Tr}\Big({Y_\nu  Y_{\nu}^{\dagger}}\Big) 
 -\frac{8}{5} g_{1}^{2} y_{t}^{4} 
 -32 g_{3}^{2} y_{t}^{4} -3 \lambda y_{t}^{4} 
 - \lambda \mbox{Tr}\Big({Y_\nu  Y_{\nu}^{\dagger}  Y_\nu  Y_{\nu}^{\dagger}}\Big)  \nonumber \\ 
 &+30 y_{t}^{6} +10 \mbox{Tr}\Big({Y_\nu  Y_{\nu}^{\dagger}  Y_\nu  Y_{\nu}^{\dagger}  Y_\nu  Y_{\nu}^{\dagger}}\Big) \, .
\end{align}

%%%%%%%%%%%%%%%%%%%%%%%%%%%%%%%%%%%%%%%%%%%%%%%%%%%%%%%%%%%%%
\subsection{Yukawa Couplings}
\label{app:seesaw-yuk}
%%%%%%%%%%%%%%%%%%%%%%%%%%%%%%%%%%%%%%%%%%%%%%%%%%%%%%%%%%%%%%%%

The one-loop and two-loop RGEs corrections to $Y_\nu$ in the (3,n) seesaw are given by

\begin{align} 
\beta_{Y_\nu}^{(1)} &  =  
\frac{3}{2} {Y_\nu  Y_{\nu}^{\dagger}  Y_\nu} 
 +Y_\nu \Big(3 y_{t}^{2}  -\frac{9}{20} g_{1}^{2}  -\frac{9}{4} g_{2}^{2} + \mbox{Tr}\Big({Y_\nu  Y_{\nu}^{\dagger}}\Big)\Big) \, ,
 \end{align}
\begin{align}  
\beta_{Y_\nu}^{(2)} & =  
\frac{1}{80} \Big(279 g_{1}^{2} {Y_\nu  Y_{\nu}^{\dagger}  Y_\nu} +675 g_{2}^{2} {Y_\nu  Y_{\nu}^{\dagger}  Y_\nu} -960 \lambda {Y_\nu  Y_{\nu}^{\dagger}  Y_\nu} 
 +120 {Y_\nu  Y_{\nu}^{\dagger}  Y_\nu  Y_{\nu}^{\dagger}  Y_\nu}  
 -540 {Y_\nu  Y_{\nu}^{\dagger}  Y_\nu} y_{t}^{2} \nonumber \\
 &-180 {Y_\nu  Y_{\nu}^{\dagger}  Y_\nu} \mbox{Tr}\Big({Y_\nu  Y_{\nu}^{\dagger}}\Big) 
 +2 Y_\nu \Big(21 g_{1}^{4} -54 g_{1}^{2} g_{2}^{2} -230 g_{2}^{4} +240 \lambda^{2}   
 +85 g_{1}^{2} y_{t}^{2} +225 g_{2}^{2} y_{t}^{2} \nonumber \\
 &+800 g_{3}^{2} y_{t}^{2} +15 g_{1}^{2} \mbox{Tr}\Big({Y_\nu  Y_{\nu}^{\dagger}}\Big) 
 +75 g_{2}^{2} \mbox{Tr}\Big({Y_\nu  Y_{\nu}^{\dagger}}\Big) 
  -270 y_{t}^{4} -90 \mbox{Tr}\Big({Y_\nu  Y_{\nu}^{\dagger}  Y_\nu  Y_{\nu}^{\dagger}}\Big) \Big)\Big) \, .
 \end{align}\\
 
 The RGEs corrections to the top-Yukawa coupling $y_t$ are given by
 
\begin{align}
\beta_{y_{t}}^{(1)} & =  
\frac{3}{2}  y_{t}^{3} 
 +y_{t} \Big( 3 y_{t}^{2}  -8 g_{3}^{2}  -\frac{17}{20} g_{1}^{2}  -\frac{9}{4} g_{2}^{2}   + \mbox{Tr}\Big({Y_\nu  Y_{\nu}^{\dagger}}\Big)\Big) \, ,
 \end{align}
 \begin{align}
\beta_{y_{t}}^{(2)} & =  
+\frac{1}{80} \Big(120 y_{t}^{5} 
 +y_{t}^{3} \Big(1280 g_{3}^{2} -180 \mbox{Tr}\Big({Y_\nu  Y_{\nu}^{\dagger}}\Big)  + 223 g_{1}^{2} -540 y_{t}^{2}  + 675 g_{2}^{2}  -960 \lambda \Big)\Big)\nonumber \\ 
 &+y_{t} \Big(\frac{1187}{600} g_{1}^{4} -\frac{9}{20} g_{1}^{2} g_{2}^{2} -\frac{23}{4} g_{2}^{4} +\frac{19}{15} g_{1}^{2} g_{3}^{2} +9 g_{2}^{2} g_{3}^{2} -108 g_{3}^{4} +6 \lambda^{2} 
  +\frac{17}{8} g_{1}^{2} y_{t}^{2} +\frac{45}{8} g_{2}^{2} y_{t}^{2} \nonumber \\ 
 &+20 g_{3}^{2} y_{t}^{2} +\frac{3}{8} g_{1}^{2} \mbox{Tr}\Big({Y_\nu  Y_{\nu}^{\dagger}}\Big) +\frac{15}{8} g_{2}^{2} \mbox{Tr}\Big({Y_\nu  Y_{\nu}^{\dagger}}\Big) 
 -\frac{27}{4} y_{t}^{4} 
 -\frac{9}{4} \mbox{Tr}\Big({Y_\nu  Y_{\nu}^{\dagger}  Y_\nu  Y_{\nu}^{\dagger}}\Big) \Big) 
 \, .
\end{align}

 %%%%%%%%%%%%%%%%%%%%%%%%%%%%%%%%%%%%%%%%%%%%%%%%%%
\section{RGEs: Type I seesaw with majoron}
\label{app:maj}
%%%%%%%%%%%%%%%%%%%%%%%%%%%%%%%%%%%%%%%%%%%%%%%%%%%

\subsection{Quartic scalar couplings}
\label{app:maj-quartic}
%%%%%%%%%%%%%%%%%%%%%%%%%%%%%%%%%%%%%%%%

The scalar sector of the majoron model is given in Eq.~\eqref{eq:maj-pot}.  It contains three scalar quartic couplings $\lambda_H, \lambda_{H\sigma}, \lambda_\sigma$ whose one-loop and two-loop RGEs are given by 

 \begin{align} 
\beta_{\lambda_H}^{(1)} & =  
+\frac{27}{200} g_{1}^{4} +\frac{9}{20} g_{1}^{2} g_{2}^{2} +\frac{9}{8} g_{2}^{4} +\lambda_{H\sigma}^{2}-\frac{9}{5} g_{1}^{2} \lambda_H -9 g_{2}^{2} \lambda_H +24 \lambda_H^{2} 
 +12 \lambda_H y_{t}^{2} +4 \lambda_H \mbox{Tr}\Big({Y_\nu  Y_{\nu}^{\dagger}}\Big)  -6 y_{t}^{4} \nonumber \\ 
 &-2 \mbox{Tr}\Big({Y_\nu  Y_{\nu}^{\dagger}  Y_\nu  Y_{\nu}^{\dagger}}\Big) \, ,
 \end{align}
 \begin{align}
\beta_{\lambda_H}^{(2)} & =  
-\frac{3411}{2000} g_{1}^{6} -\frac{1677}{400} g_{1}^{4} g_{2}^{2} -\frac{289}{80} g_{1}^{2} g_{2}^{4} +\frac{305}{16} g_{2}^{6} -4 \lambda_{H\sigma}^{3} +\frac{1887}{200} g_{1}^{4} \lambda_H 
+\frac{117}{20} g_{1}^{2} g_{2}^{2} \lambda_H -\frac{73}{8} g_{2}^{4} \lambda_H \nonumber \\
&-10 \lambda_{H\sigma}^{2} \lambda_H 
 +\frac{108}{5} g_{1}^{2} \lambda_H^{2} +108 g_{2}^{2} \lambda_H^{2} -312 \lambda_H^{3}  
  - \lambda_{H\sigma}^{2} \mbox{Tr}\Big({Y_R  Y_R^*}\Big) 
 -\frac{171}{100} g_{1}^{4} y_{t}^{2} +\frac{63}{10} g_{1}^{2} g_{2}^{2} y_{t}^{2} \nonumber \\
 &-\frac{9}{4} g_{2}^{4} y_{t}^{2} +\frac{17}{2} g_{1}^{2} \lambda_H y_{t}^{2} 
 +\frac{45}{2} g_{2}^{2} \lambda_H y_{t}^{2} +80 g_{3}^{2} \lambda_H y_{t}^{2} -144 \lambda_H^{2} y_{t}^{2} -\frac{9}{100} g_{1}^{4} \mbox{Tr}\Big({Y_\nu  Y_{\nu}^{\dagger}}\Big) 
 -\frac{3}{10} g_{1}^{2} g_{2}^{2} \mbox{Tr}\Big({Y_\nu  Y_{\nu}^{\dagger}}\Big) \nonumber \\
 &-\frac{3}{4} g_{2}^{4} \mbox{Tr}\Big({Y_\nu  Y_{\nu}^{\dagger}}\Big) +\frac{3}{2} g_{1}^{2} \lambda_H \mbox{Tr}\Big({Y_\nu  Y_{\nu}^{\dagger}}\Big) 
 +\frac{15}{2} g_{2}^{2} \lambda_H \mbox{Tr}\Big({Y_\nu  Y_{\nu}^{\dagger}}\Big) 
 -48 \lambda_H^{2} \mbox{Tr}\Big({Y_\nu  Y_{\nu}^{\dagger}}\Big)  
 -3 \lambda_H \mbox{Tr}\Big({Y_R  Y_{\nu}^{\dagger}  Y_\nu  Y_R^*}\Big) \nonumber \\
 &-\frac{8}{5} g_{1}^{2} y_{t}^{4} -32 g_{3}^{2} y_{t}^{4} -3 \lambda_H y_{t}^{4} 
 - \lambda_H \mbox{Tr}\Big({Y_\nu  Y_{\nu}^{\dagger}  Y_\nu  Y_{\nu}^{\dagger}}\Big)   
 +2 \mbox{Tr}\Big({Y_R  Y_{\nu}^{\dagger}  Y_\nu  Y_{\nu}^{\dagger}  Y_\nu  Y_R^*}\Big) 
 +2 \mbox{Tr}\Big({Y_R  Y_{\nu}^{\dagger}  Y_\nu  Y_R^*  Y_{\nu}^{T}  Y_\nu^*}\Big) \nonumber \\
 &+30 y_{t}^{6} +10 \mbox{Tr}\Big({Y_\nu  Y_{\nu}^{\dagger}  Y_\nu  Y_{\nu}^{\dagger}  Y_\nu  Y_{\nu}^{\dagger}}\Big) \, .
 \end{align}
 
 \begin{align}
\beta_{\lambda_{H\sigma}}^{(1)} & =  
-\frac{9}{10} g_{1}^{2} \lambda_{H\sigma} -\frac{9}{2} g_{2}^{2} \lambda_{H\sigma} +4 \lambda_{H\sigma}^{2} +8 \lambda_{H\sigma} \lambda_{\sigma} +12 \lambda_{H\sigma} \lambda_H 
 +\lambda_{H\sigma} \mbox{Tr}\Big({Y_R  Y_R^*}\Big)  +2 \lambda_{H\sigma} \mbox{Tr}\Big({Y_\nu  Y_{\nu}^{\dagger}}\Big) \nonumber \\
 &+6 \lambda_{H\sigma} y_{t}^{2} -4 \mbox{Tr}\Big({Y_R  Y_{\nu}^{\dagger}  Y_\nu  Y_R^*}\Big) 
 \, ,
 \end{align} 
\begin{align} 
\beta_{\lambda_{H\sigma}}^{(2)} & =  
+\frac{1671}{400} g_{1}^{4} \lambda_{H\sigma} +\frac{9}{8} g_{1}^{2} g_{2}^{2} \lambda_{H\sigma} -\frac{145}{16} g_{2}^{4} \lambda_{H\sigma}
+\frac{3}{5} g_{1}^{2} \lambda_{H\sigma}^{2} +3 g_{2}^{2} \lambda_{H\sigma}^{2} -11 \lambda_{H\sigma}^{3} -48 \lambda_{H\sigma}^{2} \lambda_{\sigma} 
- 40 \lambda_{H\sigma} \lambda_{\sigma}^{2} \nonumber \\ 
 & + \frac{72}{5} g_{1}^{2} \lambda_{H\sigma} \lambda_H +72 g_{2}^{2} \lambda_{H\sigma} \lambda -72 \lambda_{H\sigma}^{2} \lambda_H -60 \lambda_{H\sigma} \lambda_H^{2} 
 -2 \lambda_{H\sigma}^{2} \mbox{Tr}\Big({Y_R  Y_R^*}\Big) -8 \lambda_{H\sigma} \lambda_{\sigma} \mbox{Tr}\Big({Y_R  Y_R^*}\Big) \nonumber \\ 
 &+\frac{17}{4} g_{1}^{2} \lambda_{H\sigma} y_{t}^{2} +\frac{45}{4} g_{2}^{2} \lambda_{H\sigma} y_{t}^{2} +40 g_{3}^{2} \lambda_{H\sigma} y_{t}^{2} -12 \lambda_{H\sigma}^{2} y_{t}^{2}  
 -72 \lambda_{H\sigma} \lambda_H y_{t}^{2} +\frac{3}{4} g_{1}^{2} \lambda_{H\sigma} \mbox{Tr}\Big({Y_\nu  Y_{\nu}^{\dagger}}\Big) \nonumber \\
 &+\frac{15}{4} g_{2}^{2} \lambda_{H\sigma} \mbox{Tr}\Big({Y_\nu  Y_{\nu}^{\dagger}}\Big) -4 \lambda_{H\sigma}^{2} \mbox{Tr}\Big({Y_\nu  Y_{\nu}^{\dagger}}\Big) 
 -24 \lambda_{H\sigma} \lambda_H \mbox{Tr}\Big({Y_\nu  Y_{\nu}^{\dagger}}\Big) 
  +\frac{7}{2} \lambda_{H\sigma} \mbox{Tr}\Big({Y_R  Y_{\nu}^{\dagger}  Y_\nu  Y_R^*}\Big) \nonumber \\
 &-\frac{3}{2} \lambda_{H\sigma} \mbox{Tr}\Big({Y_R  Y_R^*  Y_R  Y_R^*}\Big) -\frac{27}{2} \lambda_{H\sigma} y_{t}^{4} 
 -\frac{9}{2} \lambda_{H\sigma} \mbox{Tr}\Big({Y_\nu  Y_{\nu}^{\dagger}  Y_\nu  Y_{\nu}^{\dagger}}\Big)
 +14 \mbox{Tr}\Big({Y_R  Y_{\nu}^{\dagger}  Y_\nu  Y_{\nu}^{\dagger}  Y_\nu  Y_R^*}\Big) \nonumber \\ 
 &+\frac{41}{8} \mbox{Tr}\Big({Y_R  Y_{\nu}^{\dagger}  Y_\nu  Y_R^*  Y_R  Y_R^*}\Big) +8 \mbox{Tr}\Big({Y_R  Y_{\nu}^{\dagger}  Y_\nu  Y_R^*  Y_{\nu}^{T}  Y_\nu^*}\Big)
 +\frac{39}{8} \mbox{Tr}\Big({Y_R  Y_R^*  Y_R  Y_{\nu}^{\dagger}  Y_\nu  Y_R^*}\Big) \, .
 \end{align}

 \begin{align}
\beta_{\lambda_{\sigma}}^{(1)} & =  
20 \lambda_{\sigma}^{2}  + 2 \lambda_{H\sigma}^{2}  + 2 \lambda_{\sigma} \mbox{Tr}\Big({Y_R  Y_R^*}\Big)  - \mbox{Tr}\Big({Y_R  Y_R^*  Y_R  Y_R^*}\Big) \, ,
 \end{align}
 \begin{align}
\beta_{\lambda_{\sigma}}^{(2)} & =  
+\frac{12}{5} g_{1}^{2} \lambda_{H\sigma}^{2} +12 g_{2}^{2} \lambda_{H\sigma}^{2} -8 \lambda_{H\sigma}^{3} -20 \lambda_{H\sigma}^{2} \lambda_{\sigma} -240 \lambda_{\sigma}^{3} 
 -20 \lambda_{\sigma}^{2} \mbox{Tr}\Big({Y_R  Y_R^*}\Big) -12 \lambda_{H\sigma}^{2} y_{t}^{2} \nonumber \\
 &-4 \lambda_{H\sigma}^{2} \mbox{Tr}\Big({Y_\nu  Y_{\nu}^{\dagger}}\Big) -6 \lambda_{\sigma} \mbox{Tr}\Big({Y_R  Y_{\nu}^{\dagger}  Y_\nu  Y_R^*}\Big)
 +\lambda_{\sigma} \mbox{Tr}\Big({Y_R  Y_R^*  Y_R  Y_R^*}\Big) +\frac{13}{4} \mbox{Tr}\Big({Y_R  Y_{\nu}^{\dagger}  Y_\nu  Y_R^*  Y_R  Y_R^*}\Big) \nonumber \\
 &+\frac{3}{4} \mbox{Tr}\Big({Y_R  Y_R^*  Y_R  Y_{\nu}^{\dagger}  Y_\nu  Y_R^*}\Big) +4 \mbox{Tr}\Big({Y_R  Y_R^*  Y_R  Y_R^*  Y_R  Y_R^*}\Big) \, .
\end{align}

%%%%%%%%%%%%%%%%%%%%%%%%%%%%%%%%%%%%%%%%%%%%%%%%%%%%%%%%%%%%%%%%%%%%%%%
\subsection{Yukawa Couplings}
\label{app:maj-yuk}
%%%%%%%%%%%%%%%%%%%%%%%%%%%%%%%%%%%%%%%%%%%%%%%%%%%%%%%%%%%%%%%%%%%%%%%%

The one-loop and two-loop RGEs of the Yukawa couplings $Y_\nu$, $y_t$ and $Y_R$ are given by
\begin{align} 
\beta_{Y_\nu}^{(1)} & =  
+\frac{1}{2} \Big( 3 {Y_\nu  Y_{\nu}^{\dagger}  Y_\nu}  + {Y_\nu  Y_R^*  Y_R}\Big)
 +Y_\nu \Big( 3 y_{t}^{2}  -\frac{9}{20} g_{1}^{2}  -\frac{9}{4} g_{2}^{2}  + \mbox{Tr}\Big({Y_\nu  Y_{\nu}^{\dagger}}\Big)\Big)\, , 
 \end{align}
 \begin{align} 
\beta_{Y_\nu}^{(2)} & =  
\frac{1}{80} \Big(279 g_{1}^{2} {Y_\nu  Y_{\nu}^{\dagger}  Y_\nu} +675 g_{2}^{2} {Y_\nu  Y_{\nu}^{\dagger}  Y_\nu} -960 \lambda_H {Y_\nu  Y_{\nu}^{\dagger}  Y_\nu} 
-160 \lambda_{H\sigma} {Y_\nu  Y_R^*  Y_R} 
  +120 {Y_\nu  Y_{\nu}^{\dagger}  Y_\nu  Y_{\nu}^{\dagger}  Y_\nu} \nonumber \\ 
 &-10 {Y_\nu  Y_R^*  Y_R  Y_{\nu}^{\dagger}  Y_\nu} -10 {Y_\nu  Y_R^*  Y_R  Y_R^*  Y_R} +140 {Y_\nu  Y_R^*  Y_{\nu}^{T}  Y_\nu^*  Y_R}  
 -30 {Y_\nu  Y_R^*  Y_R} \mbox{Tr}\Big({Y_R  Y_R^*}\Big) \nonumber \\
 &-540 {Y_\nu  Y_{\nu}^{\dagger}  Y_\nu} y_{t}^{2} 
 -180 {Y_\nu  Y_{\nu}^{\dagger}  Y_\nu} \mbox{Tr}\Big({Y_\nu  Y_{\nu}^{\dagger}}\Big) 
 +2 Y_\nu \Big(21 g_{1}^{4} -54 g_{1}^{2} g_{2}^{2} -230 g_{2}^{4} +20 \lambda_{H\sigma}^{2} +240 \lambda_H^{2}  \nonumber \\ 
 & +85 g_{1}^{2} y_{t}^{2} +225 g_{2}^{2} y_{t}^{2} +800 g_{3}^{2} y_{t}^{2} 
 +15 g_{1}^{2} \mbox{Tr}\Big({Y_\nu  Y_{\nu}^{\dagger}}\Big) +75 g_{2}^{2} \mbox{Tr}\Big({Y_\nu  Y_{\nu}^{\dagger}}\Big)  
 -30 \mbox{Tr}\Big({Y_R  Y_{\nu}^{\dagger}  Y_\nu  Y_R^*}\Big) \nonumber \\
 &-270 y_{t}^{4} -90 \mbox{Tr}\Big({Y_\nu  Y_{\nu}^{\dagger}  Y_\nu  Y_{\nu}^{\dagger}}\Big) \Big)\Big) \, ,
 \end{align}
 \begin{align}
\beta_{y_t}^{(1)} & =  
\frac{3}{2} y_{t}^{3}
 +y_{t} \Big(3 y_{t}^{2}  -8 g_{3}^{2}  -\frac{17}{20} g_{1}^{2}  -\frac{9}{4} g_{2}^{2}  + \mbox{Tr}\Big({Y_\nu  Y_{\nu}^{\dagger}}\Big)\Big)\, ,
\end{align}
 \begin{align} 
\beta_{y_t}^{(2)} & =  
+\frac{1}{80} \Big( 120 y_{t}^{5}  
 +y_{t}^{3} \Big(1280 g_{3}^{2}  -180 \mbox{Tr}\Big({Y_\nu  Y_{\nu}^{\dagger}}\Big)  + 223 g_{1}^{2}   -540 y_{t}^{2}  + 675 g_{2}^{2}  -960 \lambda_H \Big) \Big)\nonumber \\ 
 &+y_{t} \Big(\frac{1187}{600} g_{1}^{4} -\frac{9}{20} g_{1}^{2} g_{2}^{2} -\frac{23}{4} g_{2}^{4} +\frac{19}{15} g_{1}^{2} g_{3}^{2} +9 g_{2}^{2} g_{3}^{2} -108 g_{3}^{4} 
 +\frac{1}{2} \lambda_{H\sigma}^{2} +6 \lambda_H^{2} 
  +\frac{17}{8} g_{1}^{2} y_{t}^{2} \nonumber \\
  &+\frac{45}{8} g_{2}^{2} y_{t}^{2} 
 +20 g_{3}^{2} y_{t}^{2} +\frac{3}{8} g_{1}^{2} \mbox{Tr}\Big({Y_\nu  Y_{\nu}^{\dagger}}\Big) +\frac{15}{8} g_{2}^{2} \mbox{Tr}\Big({Y_\nu  Y_{\nu}^{\dagger}}\Big) 
  -\frac{3}{4} \mbox{Tr}\Big({Y_R  Y_{\nu}^{\dagger}  Y_\nu  Y_R^*}\Big) 
 -\frac{27}{4} y_{t}^{4} \nonumber \\
 &-\frac{9}{4} \mbox{Tr}\Big({Y_\nu  Y_{\nu}^{\dagger}  Y_\nu  Y_{\nu}^{\dagger}}\Big) \Big) \, ,
 \end{align}
 \begin{align}
 \beta_{Y_R}^{(1)} & =  
\frac{1}{2} Y_R \mbox{Tr}\Big({Y_R  Y_R^*}\Big)  + {Y_R  Y_{\nu}^{\dagger}  Y_\nu} + {Y_R  Y_R^*  Y_R} + {Y_{\nu}^{T}  Y_\nu^*  Y_R} \, ,
\end{align}
 \begin{align}
\beta_{Y_R}^{(2)} & =  
\frac{1}{40} \Big(-320 \lambda_{\sigma} {Y_R  Y_R^*  Y_R} +51 g_{1}^{2} {Y_{\nu}^{T}  Y_\nu^*  Y_R} +255 g_{2}^{2} {Y_{\nu}^{T}  Y_\nu^*  Y_R} -160 \lambda_{H\sigma} {Y_{\nu}^{T}  Y_\nu^*  Y_R} \nonumber \\ 
 & -10 {Y_R  Y_{\nu}^{\dagger}  Y_\nu  Y_{\nu}^{\dagger}  Y_\nu} -10 {Y_R  Y_{\nu}^{\dagger}  Y_\nu  Y_R^*  Y_R} +70 {Y_R  Y_R^*  Y_R  Y_R^*  Y_R} 
 -10 {Y_R  Y_R^*  Y_{\nu}^{T}  Y_\nu^*  Y_R} \nonumber \\
 &+160 {Y_{\nu}^{T}  Y_\nu^*  Y_R  Y_{\nu}^{\dagger}  Y_\nu} 
 -10 {Y_{\nu}^{T}  Y_\nu^*  Y_{\nu}^{T}  Y_\nu^*  Y_R}   
 -30 {Y_R  Y_R^*  Y_R} \mbox{Tr}\Big({Y_R  Y_R^*}\Big) -180 {Y_{\nu}^{T}  Y_\nu^*  Y_R} y_{t}^{2} \nonumber \\ 
 &+{Y_R  Y_{\nu}^{\dagger}  Y_\nu} \Big(-160 \lambda_{H\sigma}    -180 y_{t}^{2}  + 255 g_{2}^{2}  + 51 g_{1}^{2} 
 -60 \mbox{Tr}\Big({Y_\nu  Y_{\nu}^{\dagger}}\Big) \Big) 
 -60 {Y_{\nu}^{T}  Y_\nu^*  Y_R} \mbox{Tr}\Big({Y_\nu  Y_{\nu}^{\dagger}}\Big) \nonumber \\
 &+10 Y_R \Big(16 \lambda_{\sigma}^{2}  -3 \mbox{Tr}\Big({Y_R  Y_R^*  Y_R  Y_R^*}\Big)  + 4 \lambda_{H\sigma}^{2}
 -6 \mbox{Tr}\Big({Y_R  Y_{\nu}^{\dagger}  Y_\nu  Y_R^*}\Big) \Big)\Big) 
 \end{align}
 %%%%%%%%%%%%%%%%%%%%%%%%%%%%%%%%%%%%%%%%%%
\bibliographystyle{utphys}
\bibliography{bibliography} 
\end{document}